\documentclass[11pt]{article}
\usepackage{jheppub}
\usepackage{amsmath,amssymb,amsfonts,graphicx,tensor}
\usepackage{subcaption}
\usepackage{graphicx}
\usepackage{enumitem}
\usepackage{braket}
\usepackage[dvipsnames]{xcolor}
\usepackage{comment}
\usepackage{soul}
\usepackage{verbatim}

\newcommand{\EM}{\mathcal{M}}
\newcommand{\tr}{~\mathrm{tr}~}
\newcommand{\dd}{\mathrm d}
\newcommand{\xvec}{\vec{x}}
\newcommand{\yvec}{\vec{y}}
\newcommand{\Vol}{\text{Vol}}
\newcommand{\Var}{\text{Var}}

\title{Holographic RG flows and wormholes from sinusoidal scalars}
\author{Pompey Leung}
\affiliation{Department of Physics and Astronomy, University of British Columbia,\\
6224 Agricultural Road, Vancouver, BC, V6T 1Z1, Canada}

\emailAdd{pompeyleung@phas.ubc.ca}

\abstract{We study semiclassical geometries induced by turning on identical sinusoidal scalar sources on two asymptotic anti-de Sitter (AdS) boundaries. 
Varying the source strength, we find that large and small wormhole solutions appear beyond a certain threshold.
Above this threshold, the large wormhole is always the dominant saddle in the two-boundary gravitational path integral; there is no regime where wormholes are subdominant.
Under the ensemble interpretation of the gravitational path integral, this implies that once wormhole solutions exist, the putative ensemble has exponentially large fluctuations relative to the mean.
This is unlike previously studied examples of wormholes sourced by inhomogeneous matter which have a region in parameter space where wormholes are subdominant, and therefore the ensemble is sharply concentrated around the mean.
Interpreting the bulk geometries as holographic renormalization group (RG) flows, we also find that the sinusoidal modulation of scalar boundary sources results in effective holographic $\beta$-functions which are multivalued, indicating exotic RG flows. 
In particular, disconnected geometries correspond to flows that return to the starting fixed point in the infrared, while wormholes describe flows to a gapped phase in the infrared.}

\begin{document}

\maketitle

\section{Introduction} \label{sec:intro}

It is an old problem that Euclidean wormholes in holography pose a conundrum for factorization of the boundary partition function \cite{Maldacena:2004rf}.\footnote{For discussions predating AdS/CFT on spacetime wormholes in the gravitational path integral, see e.g. \cite{Lavrelashvili:1987jg, Hawking:1987mz, Coleman:1988cy, Giddings:1988cx, Giddings:1988wv, Marolf:2020xie}.}
In the absence of interactions between conformal field theories (CFTs) living on two manifolds $\Sigma_1, \Sigma_2$, the CFT partition function defined on their disjoint union factorizes into two single-CFT partition functions
\begin{equation} \label{eq:CFT_factorize}
    Z_\text{CFT}[\Sigma_1 \sqcup \Sigma_2] = Z_\text{CFT}[\Sigma_1] Z_\text{CFT}[\Sigma_2].
\end{equation}
The AdS/CFT dictionary then tells us that the CFT partition functions on each side of the equality can be replaced by bulk gravitational path integrals respecting the same boundary conditions
\begin{equation} \label{eq:grav_factorize}
    Z_\text{grav}[\Sigma_1 \sqcup \Sigma_2] = Z_\text{grav}[\Sigma_1]Z_\text{grav}[\Sigma_2].
\end{equation}
Euclidean wormholes, however, are not captured by this application of the dictionary.
Geometries connecting the boundaries $\Sigma_1$ and $\Sigma_2$ via the bulk allow the two-boundary gravitational path integral to be organized into factorized and connected parts
\begin{equation} \label{eq:grav_not_factorize}
    Z_\text{grav}[\Sigma_1 \sqcup \Sigma_2] = Z_\text{grav}[\Sigma_1]Z_\text{grav}[\Sigma_2] + Z_\text{WH}[\Sigma_1 \sqcup \Sigma_2].
\end{equation}
Here, $Z_\text{WH}$ contains all contributions from bulk geometries connecting $\Sigma_1$ and $\Sigma_2$, which include wormholes and their higher-topology cousins.
The discrepancy between Eqs.~\eqref{eq:grav_factorize} and \eqref{eq:grav_not_factorize} is the \emph{factorization problem}; Euclidean wormholes appear to spoil factorization of the CFT partition function.

One approach to resolving this conflict stems from low-dimensional models of quantum gravity.
In \cite{Saad:2019lba}, it was shown that the path integral of Jackiw--Teitelboim (JT) gravity, a fully solvable theory of quantum gravity on two-dimensional surfaces, computes the partition function for the dual theory averaged over an ensemble of random matrices.
This led to the idea that perhaps more generally, the gravitational path integral is actually calculating the ensemble-averaged CFT partition function,
\begin{equation}
    \big\langle Z_\text{CFT}[\Sigma] \big\rangle = Z_\text{grav}[\Sigma],
\end{equation}
where $\langle \cdot \rangle$ represents an average taken over some appropriate ensemble.
In fact, subsequent work provided evidence that pure AdS$_3$ gravity is dual to an ensemble of CFTs that generalizes the random matrix ensemble appearing in JT gravity \cite{Cotler:2020ugk, Cotler:2020hgz}.
By coarse-graining both CFTs at once, the ensemble averaging ``couples'' the CFTs together by statistically correlating the data that was averaged over.
This provides a plausible field-theoretic explanation for the appearance of wormholes in the gravitational path integral.

While much work has followed from this extension of the AdS/CFT dictionary to accommodate Euclidean wormholes, most, if not all, controlled calculations studying ensemble averaging so far have been carried out in theories with boundary dimension $d \leq 2$ (see e.g. \cite{Stanford:2019vob, Afkhami-Jeddi:2020ezh, Maloney:2020nni, Blommaert:2020seb, Garcia-Garcia:2020ttf}).\footnote{In higher dimensions, related work has explored alternative statistical interpretations of gravitational wormholes. See e.g. \cite{Pollack:2020gfa, Cotler:2022rud}.}
Indeed, such theories typically consist of parameters that are drawn from some natural choice of ensemble.
Notable examples include disorder averaging over coupling constants in the Sachdev--Ye--Kitaev model $(d=1)$ \cite{Maldacena:2016hyu}, and averaging over Gaussian random OPE coefficients in AdS$_3$/CFT$_2$ $(d=2)$ \cite{Belin:2020hea, Chandra:2022bqq}.
On the other hand, canonical examples of AdS/CFT with $d > 2$ where the boundary theory is e.g. $\mathcal{N}=4$ supersymmetric Yang--Mills $(d=4)$ \cite{Maldacena:1997re} or ABJM theory $(d=3)$ \cite{Aharony:2008ug} do not appear to come equipped with a suitable ensemble for averaging over \cite{Marolf:2021kjc, Schlenker:2022dyo}.
Bulk wormhole geometries with AdS asymptotics are also notoriously difficult to find in higher dimensions.
For instance, Euclidean wormholes (in particular those that can be embedded into string compactifications) have traditionally relied on the axion's ``wrong sign'' kinetic term to provide the requisite negative stress-energy supporting the wormhole throat \cite{Lavrelashvili:1987jg, Giddings:1987cg, Giddings:1989bq, Arkani-Hamed:2007cpn, Hertog:2017owm}.
Despite how well-studied these solutions are, whether or not AdS axion wormholes contribute to the gravitational path integral continues to be the subject of active investigation (see e.g. \cite{Hertog:2018kbz, Loges:2022nuw, Hertog:2024nys, Loveridge:2025dls, Marolf:2025evo, Held:2026huj, Maldacena:2026jqd}).

One promising alternative sees us return to ordinary matter; instead of an opposite-sign kinetic term, the gradient energy resulting from inhomogeneous scalar or Maxwell fields is sufficient to prop open the wormhole throat.
Indeed, \cite{Marolf:2021kjc} systematically studied numerous wormholes of this type in both top-down and bottom-up models of gravity coupled to spatially dependent matter fields.
Unfortunately, the string-theoretic wormholes are prone to brane-nucleation instabilities, and in cases where these UV-complete wormholes are free from instabilities, they may still be subdominant relative to the geometry where both asymptotic regions are disconnected from each other.
Nevertheless, stable wormholes do survive; a key result of \cite{Marolf:2021kjc} was that bottom-up models exhibit Hawking--Page-like (HP-like) phase transitions from the disconnected to connected saddle as the boundary source increases. 
These are HP-like in the sense that, of the three competing saddles (the disconnected geometry, a small wormhole, and a large wormhole), the disconnected geometry is dominant below a threshold value of the source $J_\text{HP}$, whereas the large wormhole dominates above $J_\text{HP}$.
Crucially, in the bottom-up models of \cite{Marolf:2021kjc}, there is a finite window of source strengths in which connected saddles exist but remain subdominant relative to the disconnected saddle.

We therefore arrive at a crossroads: inhomogeneous matter offers a robust way to construct stable AdS Euclidean wormholes in the bulk gravitational theory, but in the absence of microscopic duals to the bottom-up construction, their field theory interpretation remains obscure.
This work thus seeks to shed light on the boundary interpretation of higher-dimensional wormholes. 
We do this by studying a bottom-up model of AdS/CFT in which bulk scalar fields are sinusoidally modulated in the boundary directions, following the ansatz introduced in \cite{Marolf:2021kjc} for toroidal boundaries and adapted to planar boundaries in \cite{Maloney:2025tnn}.
We map the space of disconnected and connected solutions that can be constructed on two $\mathbb{R}^d$ boundaries.\footnote{Throughout this paper, we specialize to the case of $d=3$ for concreteness and comparison with \cite{Marolf:2021kjc}. Provided the appropriate holographic counterterms are derived, our methods extend to arbitrary $d$.}
With semiclassical geometries in hand, we can study the factorization properties of our Einstein-scalar model by performing a saddle-point analysis of the two-boundary gravitational path integral.
What we find is a surprising structure in the parameter space of solutions: the large wormhole dominates whenever it exists; there is \emph{no} HP-like phase transition or regime where wormholes are subdominant.
This is markedly different from the examples studied in \cite{Marolf:2021kjc}. 
Semiclassically, a wormhole-subdominant phase corresponds to an approximately self-averaging phase in a dual ensemble.
The absence of one in our model signals that the putative ensemble is strongly non-self-averaging whenever wormholes exist.
Beyond its implications for factorization, our model also admits an interesting interpretation in terms of the holographic renormalization group (RG).
Thanks to the $\mathbb{R}^d$ boundaries, we can recast our bulk geometries as single-boundary domain-wall flows \cite{Girardello:1998pd, Freedman:1999gp, deBoer:1999tgo}, revealing that the disconnected and connected solutions are dual to field theories undergoing distinct types of exotic RG flows sporting multivalued $\beta$-functions.

This paper is organized as follows: we set the scene in Section~\ref{sec:setup} by introducing the sinusoidal scalar ansatz of \cite{Marolf:2021kjc} and \cite{Maloney:2025tnn}.
In Section~\ref{sec:1bdy}, we study the one-boundary solution, i.e. the disconnected saddle, and interpret the geometry as a holographic RG flow.
In Section~\ref{sec:wormholes}, we examine the wormhole saddles and their structure in parameter space.
By ``folding'' the wormhole in half at the throat, we treat the geometry as a type of single-boundary domain-wall flow.
In Section~\ref{sec:saddles}, we compute the renormalized on-shell action for the semiclassical saddles obtained in Sections~\ref{sec:1bdy} and \ref{sec:wormholes}, and map the phase diagram for saddle dominance of the two-boundary gravitational path integral.
We close this paper with a discussion of what our results say about the factorization problem and comment on some open questions in Section~\ref{sec:discussion}.
A derivation of the holographic counterterms we use can be found in the appendix.

\section{The sinusoidal scalar ansatz} \label{sec:setup}

In asymptotically AdS geometries, certain inhomogeneous configurations of boundary sources can induce bulk geometries that are homogeneous and isotropic in the boundary directions.
The inhomogeneous wormholes of \cite{Marolf:2021kjc} are of this class and originally featured compact manifolds of non-negative curvature such as $S^3$ or $T^3$ as their boundaries. 
The same technique was adapted to find Euclidean wormholes with non-compact, $\mathbb{R}^3$ boundaries in \cite{Maloney:2025tnn}, for the purpose of constructing big-bang/big-crunch cosmologies via analytic continuation.
While more general matter configurations are possible, for concreteness we will work exclusively with Einstein gravity coupled to scalar fields throughout this paper.

Consider $d$ free complex scalar fields $\Phi_I$ coupled to gravity in a $(d+1)$-dimensional asymptotically AdS Euclidean geometry.
This is described by the Euclidean action
\begin{equation} \label{eq:bare_S}
    S = S_{\text{EH}} + S_{\text{GHY}} + S_\Phi,
\end{equation}
where each term is
\begin{align}
    S_{\text{EH}} &= -\frac{1}{2\kappa} \int_{\EM} \dd^{d+1}x \sqrt{g} \left( R - 2\Lambda \right), \\
    S_{\text{GHY}} &= - \frac{1}{\kappa} \int_{\partial \EM} \dd^dx \sqrt{h} K, \\
    S_\Phi &= \frac{1}{4\kappa} \int_{\EM} \dd^{d+1}x \sqrt{g} \sum_{I=1}^d\Big( g^{ab} \partial_a \Phi^*_I \partial_b \Phi_I + m^2 \Phi_I^*\Phi_I \Big).
\end{align}
We set $L_{\text{AdS}}=1$ and normalize the scalar action so that factors of $\kappa = 8 \pi G_N$ do not appear in the equations of motion.
 
Let us briefly sketch the ansatz for the model we will consider.\footnote{For a detailed derivation, see Section 5 of \cite{Maloney:2025tnn}.}
The equations of motion obtained from the Euclidean action Eq.~\eqref{eq:bare_S} are the Klein--Gordon equation
\begin{equation} \label{eq:KG_general}
    \left(\square - m^2 \right)\Phi_I = 0,
\end{equation}
and the Einstein equation 
\begin{equation} \label{eq:EE_general}
    G_{ab} + \Lambda g_{ab} = \kappa T_{ab}.
\end{equation}
The stress tensor due to the $d$ complex scalar fields is
\begin{equation} \label{eq:stress_general}
    T_{ab} = \frac{1}{2\kappa}\sum^d_{I=1} \left[ \partial_a \Phi^*_I \partial_b \Phi_I - \frac{1}{2}g_{ab} \Big( g^{ce} \partial_c \Phi^*_I \partial_e \Phi_I + m^2 \Phi^*_I \Phi_I \Big)\right].
\end{equation}
In contrast to the $S^3$ or $T^3$ wormholes of \cite{Marolf:2021kjc}, like \cite{Maloney:2025tnn}, we are interested in solutions with planar boundaries so that the bulk geometry in the asymptotic region is Poincar\'e--AdS.
Let $\vec{x}$ be Cartesian coordinates on the $\mathbb{R}^d$ boundary, and let $r$ be some radial coordinate normal to the boundary.
We will take the scalars to have a sinusoidal profile in the $\vec{x}$-directions so that\footnote{Note that our complex scalar is normalized to be $\sqrt{2}$ times that of \cite{Maloney:2025tnn}.}
\begin{equation} \label{eq:scalar_ansatz}
    \Phi_I(r,\vec{x}) = \varphi(r) e^{ikx^I}, \qquad I=1, \dots, d,
\end{equation}
that is, the collection of scalar fields consist of plane waves of wavenumber $k$ all orthogonal to each other, with $I$ labeling the direction of propagation on $\mathbb{R}^d$.
Since the stress tensor Eq.~\eqref{eq:stress_general} is quadratic in the scalar fields, the complex phases cancel out, leaving us with a stress tensor that is independent of $\vec{x}$.
As a consequence, the resulting solution to the Einstein equation Eq.~\eqref{eq:EE_general} will be a homogeneous and isotropic geometry.
These solutions can be expressed in Friedmann--Lema\^itre--Robertson--Walker (FLRW) form
\begin{equation} \label{eq:FLRW_metric}
    \dd s^2 = \dd \tau^2 + a(\tau)^2 \dd \vec{x}^2,
\end{equation}
where the homogeneity and isotropy of the preferred $\mathbb{R}^d$ slicing is made manifest.
The standard holographic dictionary then tells us that we can identify the coefficient of the non-normalizable mode in the near-boundary expansion of $\Phi_I$ with the source for the dual CFT operator \cite{Gubser:1998bc,Witten:1998qj}.
For example, in coordinates where $z$ is the radial direction and the asymptotic boundary lies at $z=0$, the source is defined as 
\begin{equation} \label{eq:source_def}
    J_I(\xvec) = \lim_{z \to 0} \Phi_I(z, \xvec) z^{\Delta-d}.
\end{equation}
Given our sinusoidal scalar ansatz, for us this is simply
\begin{equation} \label{eq:sinusoidal_source}
    J_I(\vec{x})=Je^{ikx^I}, \qquad I=1, \dots, d,
\end{equation}
where the free parameter $J$ is the source strength.
On the boundary theory, turning on sources $J_I$ for the bulk equations of motion is equivalent to deforming the CFT action by
\begin{equation} \label{eq:deform_CFT}
    S_\text{CFT} \longrightarrow S_\text{CFT} + \int \dd^d x ~\sum_{I=1}^d \Big(J_I(\xvec) \mathcal{O}_I^\dagger(\xvec) + J_I^*(\xvec) \mathcal{O}_I(\xvec)   \Big),
\end{equation}
where $\mathcal{O}_I$ are operators of dimension $\Delta$ dual to bulk scalar fields of mass $m^2=\Delta(\Delta-d)$.
In this way, the sources $J_I$ also act as couplings for the operators $\mathcal{O}_I$.
Note that Eq.~\eqref{eq:sinusoidal_source} is true regardless of what radial coordinates we choose to use.
This class of solutions is therefore characterized entirely by its scale factor $a(\tau)$ and the radial scalar profile $\varphi(\tau)$.

As a concrete example, our focus will be on the specific case where our boundary has dimension $d=3$, and we turn on relevant scalar operators of dimension $\Delta=2$.
The corresponding bulk scalar fields have squared mass $m^2=-2$, i.e. they are conformally coupled to Einstein gravity \cite{Marolf:2021kjc}. 
This setup represents the simplest configuration to study solutions with planar boundaries as we can avoid both conformal and matter anomaly terms when renormalizing the on-shell action.\footnote{Such anomalies appear when $d=2\mathbb{Z}$ or $\Delta = d/2+\mathbb{Z}$ respectively, and require extra logarithmic counterterms.}
While determining the appropriate holographic counterterms requires some care, our analysis below generalizes straightforwardly to other values of $d$ and $\Delta$.
For this purpose, we will leave $d$ and $\Delta$ arbitrary until it is necessary to specialize to $d=3, \Delta=2$ in our calculations.

\section{The one-boundary solution and its holographic RG flow} \label{sec:1bdy}

Let us first study the disconnected, one-boundary geometry before moving on to the connected wormhole solution.
To find a solution to the one-boundary problem, it is useful to work directly in Fefferman--Graham (FG) coordinates \cite{Fefferman:1985},
\begin{equation} \label{eq:FG_general}
    \dd s^2 = \frac{1}{z^2} \Big(\dd z^2 + \gamma_{ij}(z,x) \dd x^i \dd x^j \Big).
\end{equation}
In these coordinates, the asymptotic boundary is located at $z=0$.
Taking advantage of the homogeneity and isotropy of the sinusoidal scalar ansatz, we can greatly simplify the FG metric by setting the spatial metric to $\gamma_{ij} (z,x) = A(z) \delta_{ij}$ so that
\begin{equation} \label{eq:FG_ansatz}
    \dd s^2 = \frac{1}{z^2} \Big(\dd z^2 + A(z) \dd\vec{x}^2 \Big).
\end{equation}
We will refer to $A(z)$ as the FG scale factor to distinguish it from the FLRW scale factor $a(\tau)$ in Eq.~\eqref{eq:FLRW_metric}.

Since our solutions can always be written in FLRW form, the Einstein equation reduces to the familiar (Euclidean) Friedmann equation.
The FG coordinate analog comes from the $(z,z)$-component of Eq.~\eqref{eq:EE_general}. 
Further taking the $z$-derivative of the Friedman equation gives us a second order differential equation amenable to numerical evaluation,\footnote{This is slightly different to the second Friedmann equation as we reinsert the first Friedmann equation rather than use conservation of stress-energy after taking the $z$-derivative.}
\begin{equation} \label{eq:FG_FR3}
    \frac{A''}{A} - \frac{3}{z} \frac{A'}{A} = - \frac{1}{d-1} \left[ (d-2) \varphi'^2 + 2\frac{m^2}{z^2} \varphi^2 + \frac{k^2}{A} \varphi^2 \right], 
\end{equation}
where $(~)'\equiv \frac{\dd}{\dd z}$.
Using the sinusoidal ansatz in Eq.~\eqref{eq:scalar_ansatz} with $r=z$, the Klein--Gordon equation on the metric Eq.~\eqref{eq:FG_ansatz} takes the form 
\begin{equation} \label{eq:FG_KG}
    \varphi'' + \frac{d}{2} \frac{A'}{A} \varphi' - \frac{d-1}{z} \varphi' - \left(\frac{k^2}{A} +\frac{m^2}{z^2} \right) \varphi = 0.
\end{equation}
Eqs.~\eqref{eq:FG_FR3} and \eqref{eq:FG_KG} form the system of coupled differential equations to be solved for $A(z)$ and $\varphi(z)$.
For boundary conditions, we impose standard AdS asymptotics near the conformal boundary at $z=0$, and regularity deep in the bulk as $z \to \infty$.
From the identification of the scalar source Eq.~\eqref{eq:source_def} via the holographic dictionary, and the fact that the geometry must approach Poincar\'e--AdS as $z \to 0$, near the conformal boundary we require the following Dirichlet boundary conditions
\begin{equation} \label{eq:FG_BC1}
    \lim_{z\to 0} A(z)=1, \qquad \qquad \lim_{z \to 0} \varphi(z)z^{\Delta-d} = J.
\end{equation}
This choice corresponds to standard quantization in AdS space \cite{Klebanov:1999tb}. 
We will impose regularity deep in the bulk by demanding the vanishing of first derivatives
\begin{equation} \label{eq:FG_BC2}
    \lim_{z \to \infty} A'(z) = 0, \qquad \qquad \lim_{z\to\infty} \varphi'(z)=0.
\end{equation}

Let us briefly comment on the free parameters of our model.
Although a priori it would appear that we have two input parameters $(J, k)$, we can use the scaling symmetry of the equations of motion under $k \to \lambda k$ and $z \to \lambda^{-1}z$ to reduce this to just one free parameter.
Since the only scale invariant and hence physically meaningful combination of parameters is the dimensionless source strength $\widetilde{J} \equiv J/k^{d-\Delta}$, without loss of generality we will choose to fix $k$ and study the family of solutions by varying $J$.\footnote{The source strength scales like $J \to \lambda^{-\Delta}J$ due to how the scaling affects the power law behavior of $\varphi(z)$ near the boundary.}

\subsection{Numerical solutions} 
\label{ssec:numeric_sols_disc}
While a perturbative solution in the dimensionless source strength $\widetilde{J}$ can be found analytically, to incorporate full backreaction due to the scalar fields, we need to look for a numerical solution.
In practice, we will implement these boundary conditions numerically by evaluating the near-boundary conditions at a UV cutoff $z=\epsilon \ll1$, and the deep-bulk conditions at an IR cutoff $z=z_{\text{max}} \gg \epsilon$
\begin{alignat}{2}
    &A(\epsilon) =1, && \qquad \qquad \varphi(\epsilon) \epsilon^{\Delta-d} = J, \label{eq:IR_BC1} \\
    &A'(z_{\text{max}}) = 0, && \qquad \qquad \varphi'(z_{\text{max}}) = 0. \label{eq:IR_BC2}
\end{alignat}
We will justify why this is a reasonable approach a posteriori once we obtain the desired solutions.

To solve our boundary value problem, we will use the shooting method; swapping the boundary conditions at $z_{\text{max}}$ for two shooting parameters $A'(\epsilon)=p$ and $\varphi'(\epsilon) = q$ at $z=\epsilon$ turns it into a more tractable initial value problem.
Numerically solving the coupled differential equations Eqs.~\eqref{eq:FG_FR3} and \eqref{eq:FG_KG} with the above initial conditions gives us a two-parameter family of solutions, $A(z; p, q)$ and $\varphi(z; p, q)$.
Standard numerical root-finding can then give the desired solutions with parameters $p=p_0$ and $q=q_0$ such that $A'(z_{\text{max}}; p_0, q_0) = 0$ and $\varphi'(z_{\text{max}}; p_0, q_0) = 0$, satisfying the original boundary conditions.

\begin{figure}[h]
    \centering
    \includegraphics[width=\linewidth]{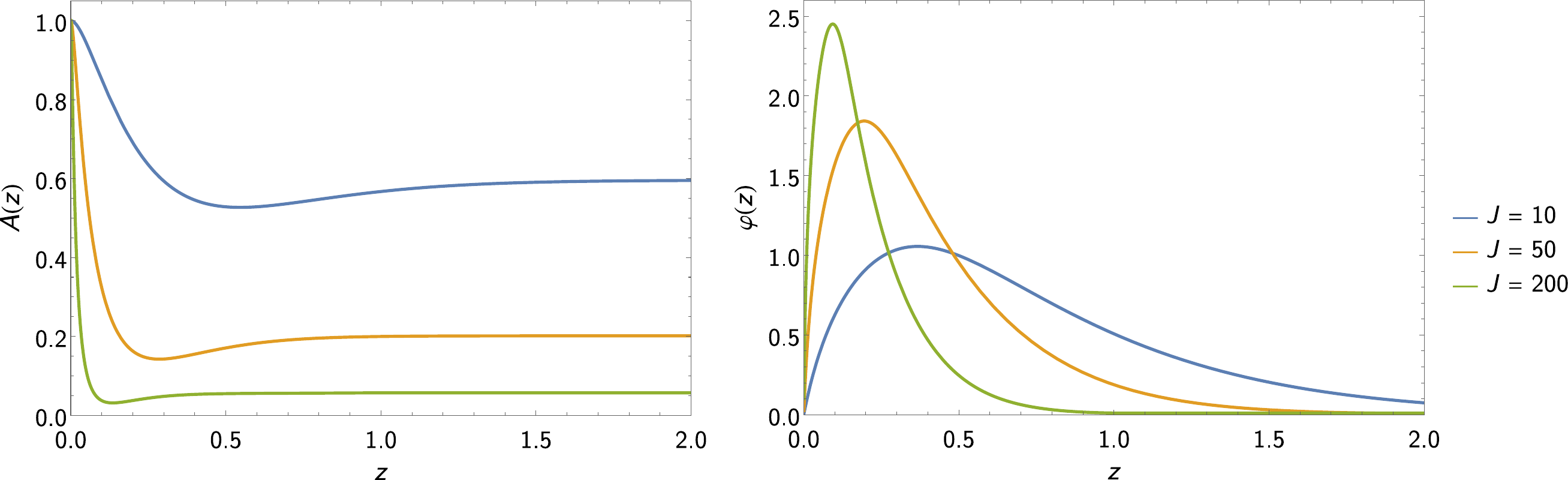}
    \caption{Fully backreacted one-boundary solutions for $d=3,~\Delta=2,~k=2$. \textit{Left.} The FG scale factor $A(z)$. \textit{Right.} The radial scalar profile $\varphi(z)$.}
    \label{fig:disconnected_sols}
\end{figure}

Examples of one-boundary solutions obtained from the shooting method are shown in Fig.~\ref{fig:disconnected_sols}.
A general feature of these solutions is that the radial scalar profile $\varphi(z)$ grows as it moves away from the boundary, reaching a maximum value before decaying to zero as it moves deeper into the bulk.
Simultaneously, the FG scale factor gets deformed away from Poincar\'{e}--AdS, with $A(z)$ decreasing accordingly as $\varphi(z)$ increases, before relaxing to the original Poincar\'{e}--AdS geometry as $\varphi(z)$ decays back to zero.\footnote{While it may appear that the minima of $A(z)$ and the maxima of $\varphi(z)$ coincide, in actuality the scalar peaks first, and the backreacting dip in the FG scale factor lags behind slightly.}
To see this last point explicitly, observe that $A(z)$ asymptotes to a value $0 < A_\infty < 1$ as $z \to \infty$.
The metric deep in the bulk then becomes
\begin{equation} \label{eq:A_infty_metric}
    \dd s^2 = \frac{1}{z^2} \Big(\dd z^2 + A_\infty \dd\vec{x}^2 \Big).
\end{equation}
This is just the Poincar\'e--AdS metric with a trivially rescaled boundary metric; the original AdS length is left unchanged.
If we denote by $z_\infty$ the value of $z$ where the scalar field has completely died down and the FG scale factor remains constant, i.e.
\begin{equation}
    A(z) = A_\infty, \qquad \varphi(z)=0, \qquad z_\infty \leq z < \infty, \label{eq:z_infty_def}
\end{equation}
we see that for $J\geq1$, the solutions return to Poincar\'{e}--AdS at $z_\infty\sim O(1)$, with $z_\infty$ decreasing as $J$ increases.
In particular, we have checked that increasing $z_{\text{max}}$ in Eq.~\eqref{eq:IR_BC2} beyond $z_\infty$ does not change the numerical solutions obtained through our shooting algorithm. 
This justifies our approach in using an IR cutoff to numerically implement regularity as $z \to \infty$.
We also expect this method to apply more generally for arbitrary boundary dimension $d$ and other relevant operators $\Delta<d$.

As Fig.~\ref{fig:disconnected_sols} suggests, we find that one-boundary solutions exist for any source strength $J$.
We will see in Section~\ref{sec:wormholes} that wormhole solutions only exist above a minimum critical value of $J$.
The solutions in Fig.~\ref{fig:disconnected_sols} were chosen in anticipation of this and for comparison reflects solutions below ($J=10$), around ($J=50$), and above ($J=200$) the critical threshold. 
At large $J$, the metric becomes strongly deformed closer to the boundary than for small $J$, but even in such cases $A(z)$ never fully reaches zero.\footnote{To give the reader a sense of the numbers involved, when $J=1000$ for example, the FG scale factor reaches a global minimum of $A(0.04) \approx 0.003$.}
We expect that the geometry only truly pinches off in the bulk when either $J \to \infty$, where the source is so strong that $A(z)$ hits zero before it has a chance to rebound, or in the infinite wavelength limit $k \to 0$, where the boundary source is just a constant and $\varphi(z)$ monotonically increases until $A(z) = 0$.
Indeed, the fact that $A_\infty$ decreases as $J$ increases is consistent with this limit where the source becomes homogeneous.

\subsection{Boomerang RG flows} 
\label{ssec:boomerang}
Asymptotically AdS spacetimes were first understood in \cite{Girardello:1998pd} and \cite{Freedman:1999gp} as the geometrization of RG flows induced by relevant deformations of the dual CFT.
Given that we have the same asymptotic geometry and similar boundary conditions, it is natural to want to interpret our one-boundary solutions through the lens of holographic RG flows.
The main difference here is that the putative flow dual to our geometry stems from sinusoidal sources on the boundary, unlike the usual case where the sources are homogeneous.

\begin{figure}[h]
    \centering
    \includegraphics[width=0.6\linewidth]{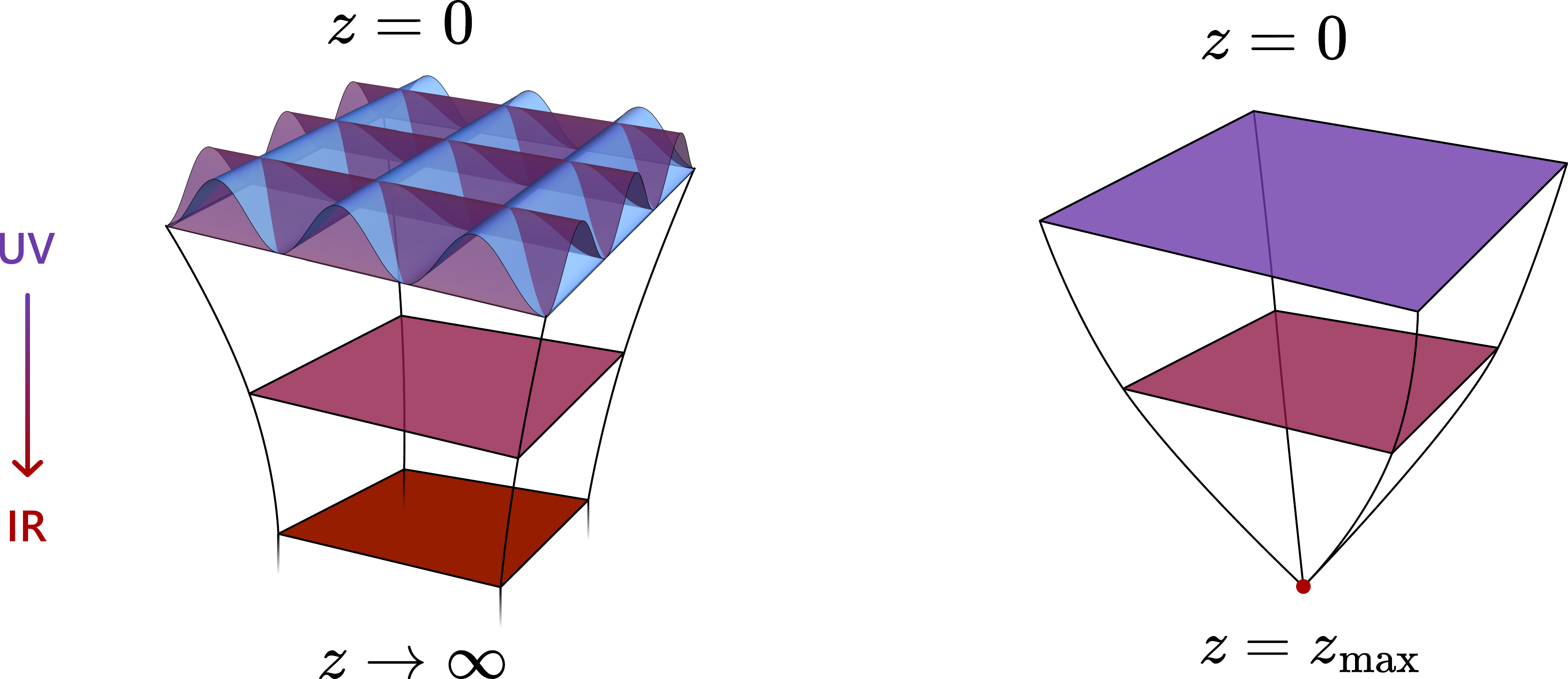}
    \caption{Schematic illustrations of bulk geometries corresponding to different holographic RG flows of a single CFT when a different IR fixed point does not exist. \textit{Left.} Sinusoidal sources induce a relevant deformation that extends infinitely deep into the bulk. \textit{Right.} Homogeneous relevant deformations can pinch off smoothly at finite $z$.}
    \label{fig:rg_geoms}
\end{figure}

To better appreciate this difference, let us briefly recall what happens for homogeneous deformations.
In the framework of the holographic renormalization group, the radial coordinate is identified with the renormalization scale as $\mu \sim 1/z$ \cite{Peet:1998wn, Porrati:1999ew, Balasubramanian:1999jd}.
The boundary CFT defined on $\mathbb{R}^d$ at $z=0$ therefore corresponds to a UV fixed point of the RG flow.
Turning on sources for scalar operators with dimension $\Delta<d$ induces a relevant deformation of the UV CFT, the physics of which is captured by the geometry away from the boundary.
A central result which extends the celebrated $c$-theorem of Zamolodchikov \cite{Zamolodchikov:1986gt} is the holographic $c$-theorem established in \cite{Freedman:1999gp}.
It guarantees that the holographically defined $c$-function must monotonically decrease under nontrivial RG flow under the assumption of Poincar\'e invariance and the null energy condition (NEC).
Additionally, it is well known that for homogeneous deformations, if the geometry cannot flow to another AdS vacuum, i.e. an IR CFT, then the geometry terminates at some finite depth $z_\text{max}$ (Fig.~\ref{fig:rg_geoms}, right), where the dual field theory is gapped or confining in the IR \cite{Girardello:1999hj, Girardello:1999bd, Polchinski:2000uf}.
In our setup where the scalar potential is quadratic so that no other AdS extrema are available, this is precisely the expectation alluded to at the end of Section~\ref{ssec:numeric_sols_disc}.

In contrast, we find that for our one-boundary solutions, relevant deformations induced by sinusoidal boundary sources are not gapped as the geometry extends smoothly to \mbox{$z \to \infty$} (Fig.~\ref{fig:rg_geoms}, left).
One way to see this explicitly is to compute the holographic $\beta$- and $c$-functions introduced in \cite{Freedman:1999gp}, which quantitatively probe how the CFT evolves under RG flow.
Standard applications of this framework compute $\beta$-functions for spatially independent couplings (sources) as a gradient flow of the corresponding homogeneous bulk scalar fields.
Since our couplings $J_I(\xvec)$ in Eq.~\eqref{eq:deform_CFT} are spatially dependent, however, we emphasize that what we calculate here is not the $\beta$-function for the fully spatially-dependent sources, but rather an \emph{effective} holographic $\beta$-function obtained from the spatially homogeneous radial mode $\varphi(z)$ common to our bulk scalar fields.\footnote{Several studies have explored aspects of a $\beta$-function for spatially-dependent couplings, but this subject remains somewhat underdeveloped. See e.g. \cite{Hoyos:2012xc, Hartnoll:2015faa, Aharony:2015aea}.}
Nevertheless, given our sinusoidal scalar ansatz, one reasonable interpretation of the resulting $\beta$-function is that it describes a spatially-dependent running coupling of the form
\begin{equation}
    g_I(\mu, \xvec) \sim \varphi(\mu) e^{ikx^I},
\end{equation}
with $J$ setting the UV value for the running coupling.

\begin{figure}[h]
    \centering
    \includegraphics[width=\linewidth]{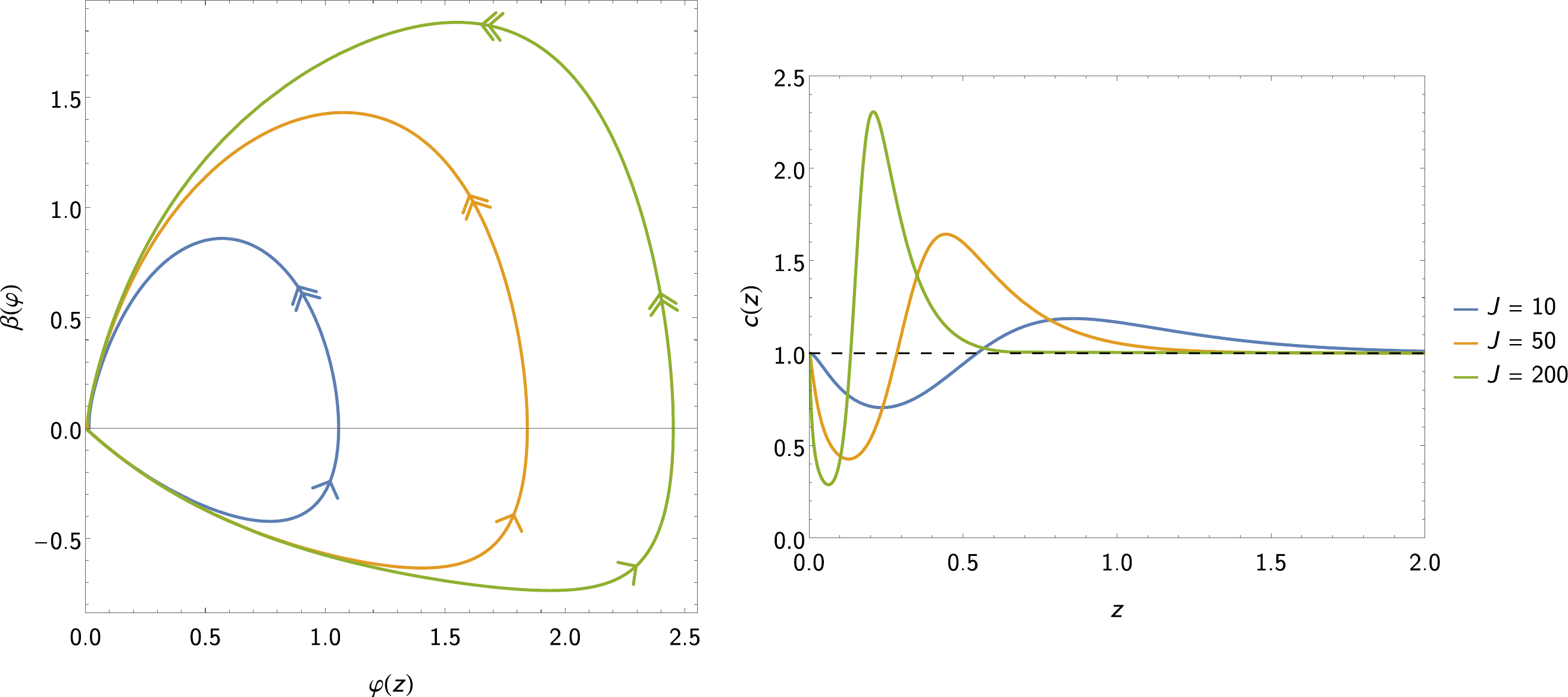}
    \caption{RG flow functions for $d=3,~\Delta=2,~k=2$ showing boomerang flows. \textit{Left.} The multivalued $\beta$-function with two branches. Arrows indicate the direction of flow along each branch: single arrows label the negative ($\beta < 0$) branches, and double arrows label the positive ($\beta > 0$) branches. Flows return to their starting fixed point as $z \to \infty$. \textit{Right.} The holographic $c$-function. $c(z)$ returns to its starting value $c_0$ indicated by the dotted black line (here normalized to 1) as $z \to \infty$.}
    \label{fig:RG_obs}
\end{figure}

Under this interpretation, the holographic $\beta$- and $c$-functions in the metric Eq.~\eqref{eq:FG_ansatz} are
\begin{equation} \label{eq:beta_c_func}
    \beta(\varphi) = -\frac{\dd \varphi(z)}{\dd \ln z}, \qquad c(z) = c_0 \left(1-\frac{z}{2}\frac{A'(z)}{A(z)} \right)^{1-d},
\end{equation}
where $-\ln z$ is the RG time.
Applying Eq.~\eqref{eq:beta_c_func} to our one-boundary solutions, we see in Fig.~\ref{fig:RG_obs} that both $\beta(\varphi)$ and $c(z)$ exhibit non-monotonic, cyclic behavior under RG flow.
Indeed, the $\beta$-function is multivalued with a $\beta(\varphi) < 0$ and a $\beta(\varphi) > 0$ branch.
For each flow, $\beta(\varphi)$ leaves the UV fixed point located at $\varphi = 0$, traversing the negative branch until it reaches a maximum value of the scalar field, $\varphi_\text{max}$.
The $\beta$-function then switches to the positive branch and flows back toward $\varphi = 0$; the fixed point where the flow originated is now an IR attractor.
If we go by usual RG lore, the zero of $\beta(\varphi)$ at $\varphi_\text{max}$ belies the fact that it is \emph{not} a fixed point of the RG transformation.
Unlike single-valued $\beta$-functions encountered in standard quantum field theories, multivalued $\beta$-functions have richer renormalization structures, and a more suitable diagnostic is the RG ``acceleration'' \cite{Curtright:2010hq} 
\begin{equation} \label{eq:RG_accel}
    -\frac{\dd \beta(\varphi)}{\dd \ln z} = \beta(\varphi) \frac{\dd \beta(\varphi)}{\dd \varphi}.
\end{equation}
We have verified that at the branch point $\varphi_\text{max}$, the RG acceleration is finite and positive.
$\varphi_\text{max}$ is therefore a genuine ``turning point'' of the RG flow where $\beta$ changes sign as it passes through; there is no scale invariant theory at the branch point.\footnote{True fixed points have all higher $\ln z$-derivatives of $\beta$ vanishing such that the flow lands sufficiently ``softly'' at the zero \cite{Curtright:2010hq}.}
In a similar vein, the holographic $c$-function initially decreases from $c_0$ as expected, but reverses course and overshoots past $c_0$, before settling down back to $c_0$ as $z\to\infty$.
The point where $c(z)$ momentarily returns to $c_0$ in the middle of the flow coincides with the point where $A(z)$ reaches its global minimum.\footnote{Incidentally, this is \emph{not} at the same point of the flow as the turning point of the $\beta$-function because $A'(z)$ and $\varphi'(z)$ reaches their respective extrema at different $z$.}
This type of flow where the theory leaves the UV repeller only to return to it as an IR attractor is suitably known as a \emph{boomerang} RG flow \cite{Donos:2017ljs, Donos:2017sba}. 

Are these one-boundary boomerang flows pathological given that they appear to violate the $c$-theorem?
Monotonicity of the holographic $c$-function requires Poincar\'e invariance and the NEC \cite{Freedman:1999gp}.
Since we work in Euclidean signature and our scalar fields are sinusoidal, neither condition holds and there is no guarantee that $c(z)$ should be monotonic.
Indeed, many known examples of consistent single-coupling RG flows that exhibit non-monotonic, chaotic, or other exotic behaviors have multivalued $\beta$-functions \cite{Curtright:2011qg, Ilderton:2020tnv, Kiritsis:2016kog}.
The non-monotonicity of the holographic $c$-function and the two-branched $\beta$-function in Fig.~\ref{fig:RG_obs} therefore indicate that our boomerang flows fall under this class of exotic RG flows that are physically sensible.
While this does mean we can no longer interpret $c(z)$ as a measure of effective degrees of freedom at intermediate scales, in the spirit of the $c$-theorem, we can nevertheless identify $c(z)$ with the central charge at fixed points of these boomerang flows.\footnote{To be precise, in the same vein as \cite{Casini:2011kv}, because our main example is in $d=3$ boundary dimensions where the trace anomaly is absent, the holographic $c$-function is expected to be proportional to the $F$-coefficient proposed by \cite{Jafferis:2011zi} which was proven to obey an $F$-theorem in \cite{Casini:2012ei}.}
The fact that the $\beta$-function returns to the origin, $c(z)$ returns to $c_0$, and the geometry deep in the bulk as given by Eq.~\eqref{eq:A_infty_metric} is conformally equivalent to the boundary geometry are three diagnostics that point towards a single fact: the UV and IR fixed points describe one and the same CFT!

To get an intuition for why the RG flow for one-boundary solutions return, we can examine the effective action for $\varphi(z)$
\begin{equation}
    S_\varphi = \frac{1}{4\kappa} \sum^{d}_{I=1} \int \dd^{d+1} x \sqrt{g} \Bigg[ \partial_z \varphi \partial^z \varphi + \underbrace{\left(\frac{k^2 z^2}{A(z)} + m^2 \right)}_{m^2_{\mathrm{eff}}(z)} \varphi^2 \Bigg].
\end{equation}
Because the bulk scalar mass is related to the dual operator dimension by $m^2 = \Delta(\Delta-d)$, from the scale-dependent effective mass term $m^2_{\mathrm{eff}}(z)$ for $\varphi$, it is clear that even if the initial deformation was relevant in the UV ($m^2_{\mathrm{eff}}(0) < 0$), it effectively becomes irrelevant in the IR ($m^2_{\mathrm{eff}}(z) > 0$).
Unlike the homogeneous case where the scalar mass remains unchanged along the RG flow and the scalar field grows monotonically, here $\varphi(z)$ begins to decay once its effective mass becomes positive, allowing the gravitational backreaction to settle, and $A(z)$ to eventually relax back to Poincar\'{e}--AdS.
It is crucial to note that our multivalued $\beta$-functions are a direct consequence of the gradient energy of the sinusoidal scalars, giving rise to both the non-monotonicity and cyclic nature of the boomerang RG flows.

\section{The planar Euclidean wormhole and its holographic RG flow} \label{sec:wormholes}
We now turn our attention to the search for wormhole solutions. 
Given two asymptotic boundaries in the gravitational path integral, this is the leading connected saddle that competes with the disconnected saddle consisting of two copies of the one-boundary solution found in Section~\ref{sec:1bdy}.
These wormhole geometries are best described using the conformal FLRW metric
\begin{equation} \label{eq:conformal_FRW}
    ds^2 = a^2(\eta) (d\eta^2 + d\vec{x}^2),
\end{equation}
where the asymptotic boundaries are located at finite $\eta$.
Since we do not pursue perturbative solutions here, this property makes $(\eta, \vec{x})$ a convenient choice of coordinates for numerics as the entire solution exists within a finite domain.\footnote{This is especially useful when we evaluate the action integral for this saddle in Section~\ref{sec:saddles}, as we can do so without truncating the domain of integration. For the disconnected saddle we will resort to using an IR cutoff at $z=z_{\text{max}}$ to numerically approximate the upper bound at $z=\infty$.}

\begin{figure}[h]
    \centering
    \includegraphics[width=\linewidth]{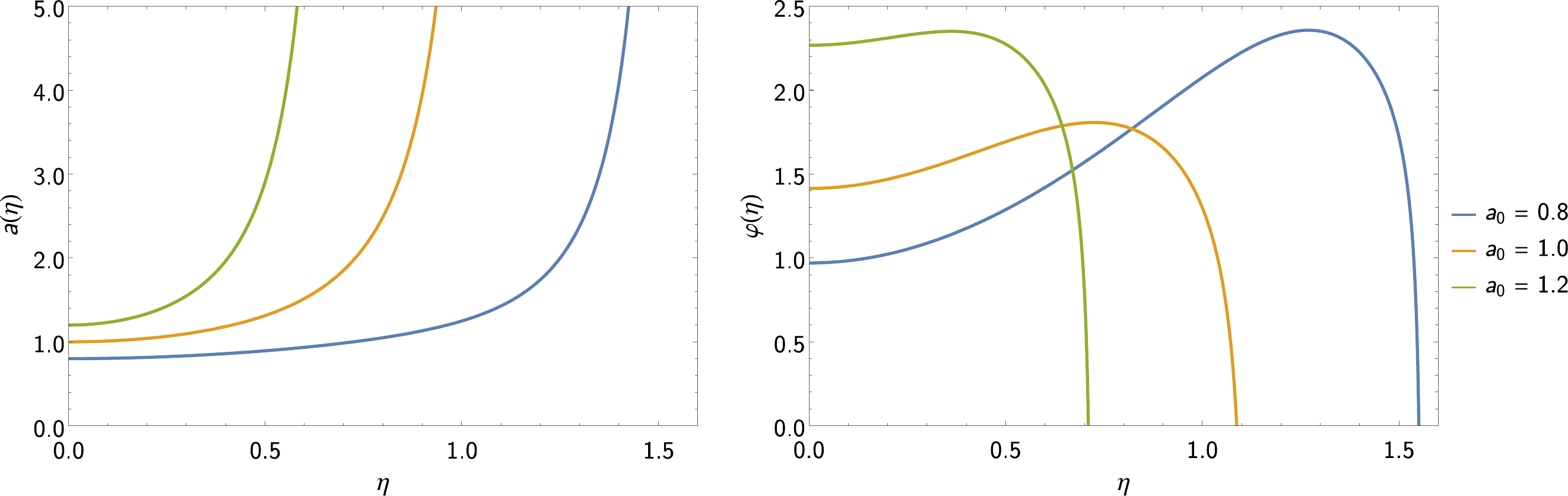}
    \caption{$\mathbb{Z}_2$-symmetric wormhole solutions for $d=3, \Delta=2, k=2$. Only the region from the throat $(\eta = 0)$ to the right boundary $(\eta = +\eta_*)$ is shown. \textit{Left.} The FLRW scale factor $a(\eta)$. \textit{Right.} The radial scalar profile $\varphi(\eta)$. The asymptotic boundary at $\eta = \eta_*$ is where $\varphi(\eta)$ reaches zero.}
    \label{fig:connected_sols}
\end{figure}

The equations of motion for conformal FLRW coordinates can be obtained in the same manner as for the one-boundary problem above.
The first and second order Friedmann equations come from the $\eta\eta$-component of the Einstein equation, Eq.~\eqref{eq:EE_general}, and its $\eta$-derivative respectively
\begin{align}
    \frac{\dot{a}^2}{a^2} &= \frac{1}{2(d-1)} \Big[ \dot{\varphi}^2 - \left( m^2 a^2 + k^2 \right) \varphi^2 \Big] + a^2, \label{eq:FRW_FR1} \\
    \frac{\ddot{a}}{a} - \frac{\dot{a}^2}{a^2} &= -\frac{1}{2(d-1)} \Big[ (d-1)\dot{\varphi}^2 + m^2 a^2 \varphi^2 \Big] + a^2, \label{eq:FRW_FR2}
\end{align}
where $\dot{(~~)}\equiv \frac{\dd}{\dd \eta}$.
Plugging Eq.~\eqref{eq:FRW_FR1} into Eq.~\eqref{eq:FRW_FR2} to remove the $\dot{a}$ term, we arrive at
\begin{equation} \label{eq:FRW_FR3}
    \frac{\ddot{a}}{a} = -\frac{1}{2(d-1)} \Big[ (d-2) \dot{\varphi}^2 + (2m^2 a^2 + k^2)\varphi^2\Big] + 2a^2.
\end{equation}
As before, assuming the sinusoidal ansatz Eq.~\eqref{eq:scalar_ansatz} with $r=\eta$, the Klein--Gordon equation in the metric Eq.~\eqref{eq:conformal_FRW} takes the form
\begin{equation} \label{eq:FRW_KG}
    \ddot{\varphi} + (d-1)\frac{\dot{a}}{a} \dot{\varphi} - (m^2 a^2 + k^2) \varphi = 0.
\end{equation}
The goal here is to solve the coupled system Eqs.~\eqref{eq:FRW_FR3} and \eqref{eq:FRW_KG} for $a(\eta)$ and $\varphi(\eta)$.
We look for the simplest wormhole where the geometry is $\mathbb{Z}_2$-symmetric around $\eta=0$.
This puts the asymptotic boundaries at $\eta = \pm \eta_*$ and is equivalent to turning on identical sources on both boundaries, rather than allowing them to independently differ.
Unlike the one-boundary case where we had a boundary value problem, here we can find wormhole solutions by imposing initial conditions at $\eta=0$ that guarantee $\mathbb{Z}_2$ symmetry, and then numerically solve the equations of motions between $0 \leq \eta \leq \eta_*$.
In other words, we want solutions that are even functions in $\eta$, satisfying
\begin{alignat}{2}
    &a(0) = a_0, && \qquad \qquad \varphi(0) = \sqrt{\frac{2(d-1)}{m^2 + k^2/a_0^2}} \equiv \varphi_0, \label{eq:FRW_IC1} \\
    &\dot{a}(0) = 0, && \qquad \qquad \dot{\varphi}(0) = 0. \label{eq:FRW_IC2}
\end{alignat}
Note that the throat size $a_0$ of the wormhole is a free parameter, whereas the scalar amplitude at the throat $\varphi_0$ is constrained by Eq.~\eqref{eq:FRW_FR1}.
While the location of the boundary $\eta_*$ is unknown a priori, standard numerical integration will dynamically give us $\eta_*$ as the point where integration stops, provided we choose a large enough domain of integration for the numerical algorithm to attempt integrating over.
Examples of wormhole solutions for various throat sizes are shown in Fig.~\ref{fig:connected_sols}.

We have verified that $a(\eta)$ has simple poles at $\eta = \pm\eta_*$ so that the geometry near the boundaries is asymptotically AdS.
We see that as $a_0$ increases, $\eta_*$ moves closer to the center of the wormhole.
While $a(\eta)$ behaves in a predictable way as we increase $a_0$, note that $\varphi(\eta)$ appears to ``flatten'' no further than $a_0 \simeq 1$ before rebounding as $a_0$ increases; its global maximum does not keep decreasing.
To understand why this is, it will be instructive to see where wormhole solutions live in the space of $J$ and $a_0$.

\subsection{The parameter space structure of wormhole solutions} 
\label{ssec:param_space}

\begin{figure}[h]
    \centering
    \includegraphics[width=0.75\linewidth]{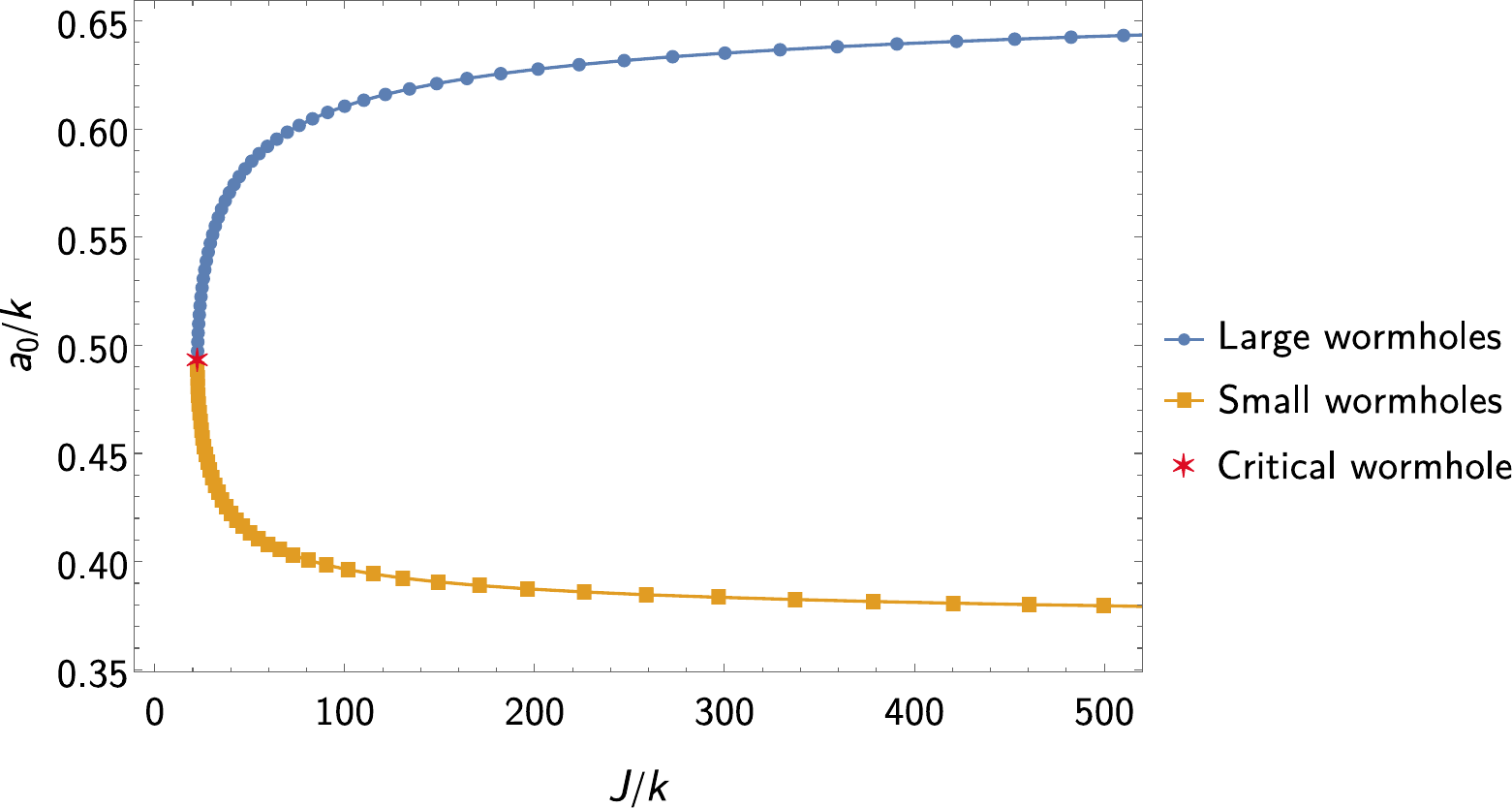}
    \caption{Small (orange squares) and large (blue disks) branches of wormhole solutions for $d=3, \Delta=2$ in the parameter space of dimensionless source strength $J/k$ against throat size $a_0/k$. Both branches coalesce into one ``critical wormhole'' (red star) at $(J/k)_\text{crit} \simeq 22.5$. No wormholes exist below this threshold.}
    \label{fig:large_vs_small}
\end{figure}

To compare the one-boundary and wormhole solutions on equal footing, we need to understand what values of $J$ give rise to each wormhole solution.
Since our wormholes were found by fixing $a_0$, we have to extract the corresponding $J$ by applying the holographic dictionary  to the asymptotics of $\varphi(\eta)$. 
Recalling the exposition surrounding Eq.~\eqref{eq:source_def}, near the boundary, the leading behavior of $\varphi(\eta)$ is given by its non-normalizable and normalizable modes, the coefficients of which correspond to the boundary source and VEV of the dual scalar operator \cite{Gubser:1998bc,Witten:1998qj}.
Changing to near-boundary FG coordinates where $z \simeq 1/a(\eta) \simeq \eta_*-\eta$, for $d=3$ and $\Delta=2$ we have\footnote{More generally, we have Eq.~\eqref{eq:phi_asymp} for any $d$ and $\Delta$ where conformal anomalies are absent.}
\begin{equation} \label{eq:extrap_coeffs_example}
    \varphi(z) \simeq Jz + \braket{\mathcal{O}}z^2 + \cdots.
\end{equation}
Extracting the source is then simply a matter of fitting $\varphi(\eta)$ in a small window around $\eta_*$ to the power law approximation above.
From the initial conditions in Eq.~\eqref{eq:FRW_IC1}, it is apparent that the family of solutions is controlled by the ratio $a_0/k$ that appears in the denominator of $\varphi_0$.
Thus, all physically distinct wormhole solutions can be characterized by $(J/k,~ a_0/k)$.
For a given throat size $a_0/k$, the corresponding dimensionless source strength $J/k$ that supports the wormhole solution is shown in Fig.~\ref{fig:large_vs_small}.

It turns out that for a given source, there are generally two connected solutions: a small and a large wormhole.
As $J/k$ decreases and reaches a critical value $(J/k)_{\text{crit}}$, both branches of wormholes merge into a single wormhole solution we dub the ``critical'' wormhole.
Below this threshold however, no wormhole solutions exist.\footnote{Presumably, the solution curve moves to somewhere in the complex $(a_0/k)$-plane. Indeed, complex saddles to the gravitational path integral arising from this type of bifurcation have been found even in the standard Hawking--Page transition \cite{Mahajan:2025bzo}. What role such complex saddles may play for wormholes and factorization is an interesting question that will be left for future work.}
In the $d=3, \Delta = 2$ example shown above in Fig.~\ref{fig:large_vs_small}, this critical wormhole occurs at $(J/k)_\text{crit} \simeq 22.5$ and has a throat size of $(a_0/k)_\text{crit}\simeq 0.493$.
The $a_0 = 1.0$ solution (orange curve) is therefore a close representation of the critical wormhole.
This bifurcating structure of wormhole saddles is in accordance with the results of \cite{Marolf:2021kjc} for various bottom-up models with compact boundary manifolds.

Unlike \cite{Marolf:2021kjc} however, our $\mathbb{R}^3$ wormholes only exist in a finite band between a minimum and maximum value of $a_0/k$.
In particular, $a_0/k$ is upper bounded by the denominator of $\varphi_0$ in Eq.~\eqref{eq:FRW_IC1} reaching zero.
For $d=3, \Delta=2$, this is $(a_0/k)_\text{max} = \sqrt{2}/2 \simeq 0.707$.
As $a_0/k$ approaches this value from below, $\varphi(0)$ goes from being local minimum to a global maximum, and the asymptotic boundary gets pushed in towards the throat at $\eta = 0$, eventually reaching it in the limit $a_0/k \to (a_0/k)_\text{max}$, or equivalently as $J/k \to \infty$ for the large branch of wormholes.
On the other hand, we have a lower bound on the throat size of $(a_0/k)_\text{min}\simeq0.373$ determined numerically.
As $a_0/k$ approaches this from above, the global maximum of $\varphi(\eta)$ gets pushed out towards the boundary at $\eta_*$, reaching it as $a_0/k \to (a_0/k)_\text{min}$, or as $J/k \to \infty$ in the small branch of wormholes.
We therefore see that in Fig.~\ref{fig:connected_sols}, the $a_0=1.0$ solution (orange curve) serves to demarcate the small and large branches of wormholes as represented by the $a_0 = 0.8$ solution (blue curve) and $a_0 = 1.2$ solution (green curve), which themselves are examples of wormholes with throat sizes close to $(a_0/k)_\text{min}$ and $(a_0/k)_\text{max}$ respectively.
Attempts to find solutions with initial conditions outside $(a_0/k)_\text{min} < a_0/k < (a_0/k)_\text{max}$ result in geometries that do not satisfy AdS boundary conditions; $a(\eta)$ goes to zero at finite $\eta$.

\subsection{Folding the wormhole to get gapped RG flows}
\label{ssec:gapped}
\begin{figure}[h]
    \centering
    \includegraphics[width=0.95\linewidth]{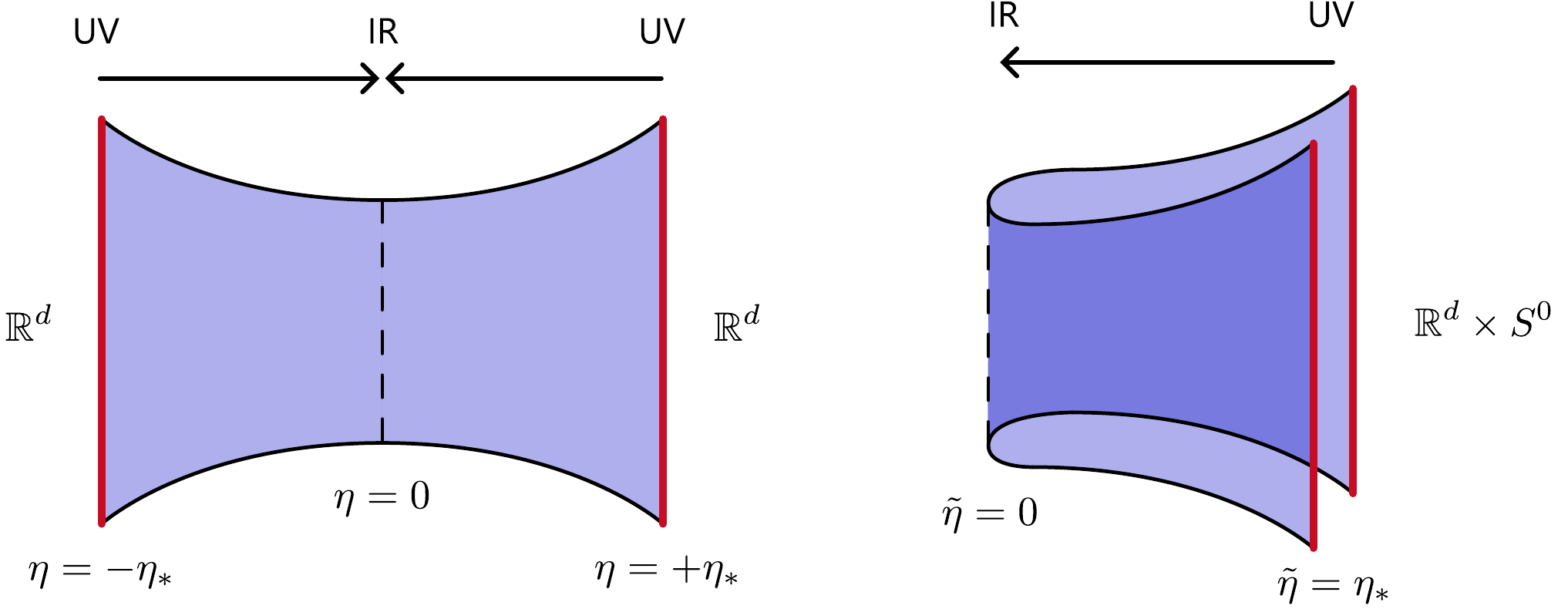}
    \caption{Equivalent interpretations of the holographic RG flow described by a $\mathbb{Z}_2$-symmetric wormhole. \textit{Left.} The two-sided RG flow where two copies of a CFT on $\mathbb{R}^d$ join up at the wormhole throat in the IR. \textit{Right.} Folding the wormhole in half and defining $\tilde{\eta} = |\eta|$ gives a one-sided RG flow: a single CFT on $\mathbb{R}^d \times S^0$ flows to a gapped theory in the IR, where the $S^0$ ``pinches-off'' at the wormhole throat.}
    \label{fig:folded_wormhole}
\end{figure}

It is natural to imagine that a wormhole describes two RG flows that start separately in the UV and join in the IR at the throat (Fig.~\ref{fig:folded_wormhole}, left).
This is certainly one of several interpretations that have been ascribed to wormholes as holographic RG flows (see \cite{Ghodsi:2022umc, Chandra:2024bqz} for discussions).
Here we adopt the perspective of \cite{VanRaamsdonk:2021qgv, Antonini:2022blk, Antonini:2022opp} and make this intuition precise.
In a similar spirit as the folding trick for boundary CFTs \cite{Wong:1994np}, we can view our $\mathbb{Z}_2$-symmetric wormholes not as a two-sided flow for two CFTs on $\mathbb{R}^d$, but instead as a \emph{one-sided} holographic RG flow for a CFT defined on $\mathbb{R}^d \times S^0$, where the flow terminates at the wormhole throat after traversing a finite radial depth (Fig.~\ref{fig:folded_wormhole}, right).
The IR theory is therefore \emph{gapped}\footnote{Since our theory is defined primarily in Euclidean signature, strictly speaking we do not have a Hamiltonian as in Lorentzian signature. Our usage of the term ``gapped'' here refers to the exponential decay of two-point correlators for large separations with a weight determined by the lowest eigenvalue of the scalar wave operator on the wormhole geometry. This plays the role analogous to that of the Lorentzian mass gap.}; the throat of the wormhole is where the $S^0$ ``pinches off'', much like the confinement mechanism of \cite{Witten:1998zw} where the radius of the $S^1$ smoothly contracts to zero.
In the following, we will study the consequences of interpreting the wormhole geometry as a gapped holographic RG flow.

\begin{figure}[h]
    \centering
    \includegraphics[width=\linewidth]{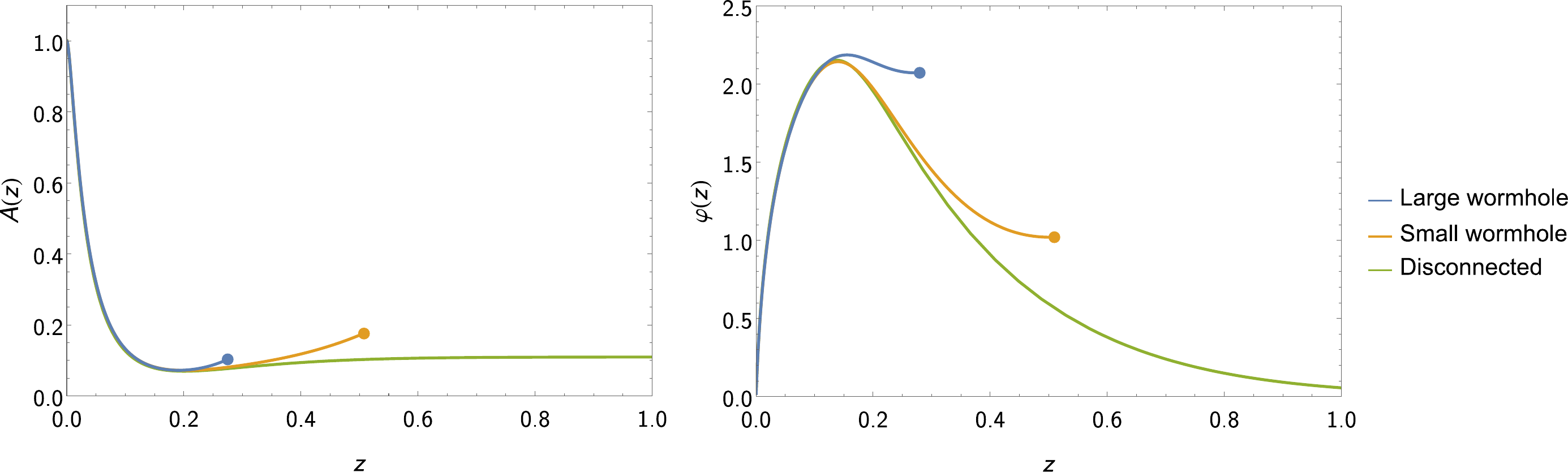}
    \caption{The large wormhole, small wormhole, and disconnected solutions in FG coordinates for $d=3,~\Delta=2,~J/k=50$. Wormhole geometries terminate at the throat ($z=z_0$) as indicated by solid circles. \textit{Left.} FG scale factor $A(z)$ \textit{Right.} Radial scalar profile $\varphi(z)$.}
    \label{fig:all_3_sols}
\end{figure}

To fold our wormhole in half, let us define $\tilde{\eta} = |\eta|$ as the radial coordinate so that the entire geometry lives between $\tilde{\eta} \in(0, \eta_*)$.
This allows us to find FG coordinates with which we can interpret the folded wormhole as a holographic RG flow.
Given a wormhole solution, Eqs.~\eqref{eq:FG_ansatz} and \eqref{eq:conformal_FRW} tell us that a one-to-one mapping from $\tilde{\eta}$ to $z$ can be found by solving
\begin{equation}
    \frac{\dd z}{\dd \tilde{\eta}} = -a(\tilde{\eta}) z,
\end{equation}
with the boundary condition
\begin{equation}
    \lim_{\tilde{\eta} \to \eta_*}a(\tilde{\eta}) = \frac{1}{\eta_*-\tilde{\eta}} = \frac{1}{z}.
\end{equation}
In practice, this amounts to numerically integrating from zero to some UV cutoff $\eta_*-\epsilon$.
The result is a one-to-one function $z(\tilde{\eta}) \in (0, z_0)$, where the boundary is now at $z=0$, and the throat is at some finite radial depth $z=z_0$.
To see how the wormhole solutions compare with the disconnected solution, we fix $J/k = 50$ as an example and show them all on the same plot in Fig.~\ref{fig:all_3_sols}.

Unsurprisingly, the near-UV flow of all three saddles coincides because they share the same boundary source.
The key difference between the connected and disconnected solutions lies in the IR: the wormhole saddles terminate at finite $z$ whereas the disconnected saddle extends infinitely deep into the bulk.
The small wormhole and disconnected geometry share the same extrema within numerical precision for $A(z)$ and $\varphi(z)$.
On the other hand, the extrema for the large wormhole are shifted relative to the other saddles; the minimum of $A(z)$ occurs at smaller $z$, and the maximum of $\varphi(z)$ occurs at larger $z$. 

\begin{figure}[h]
    \centering
    \includegraphics[width=\linewidth]{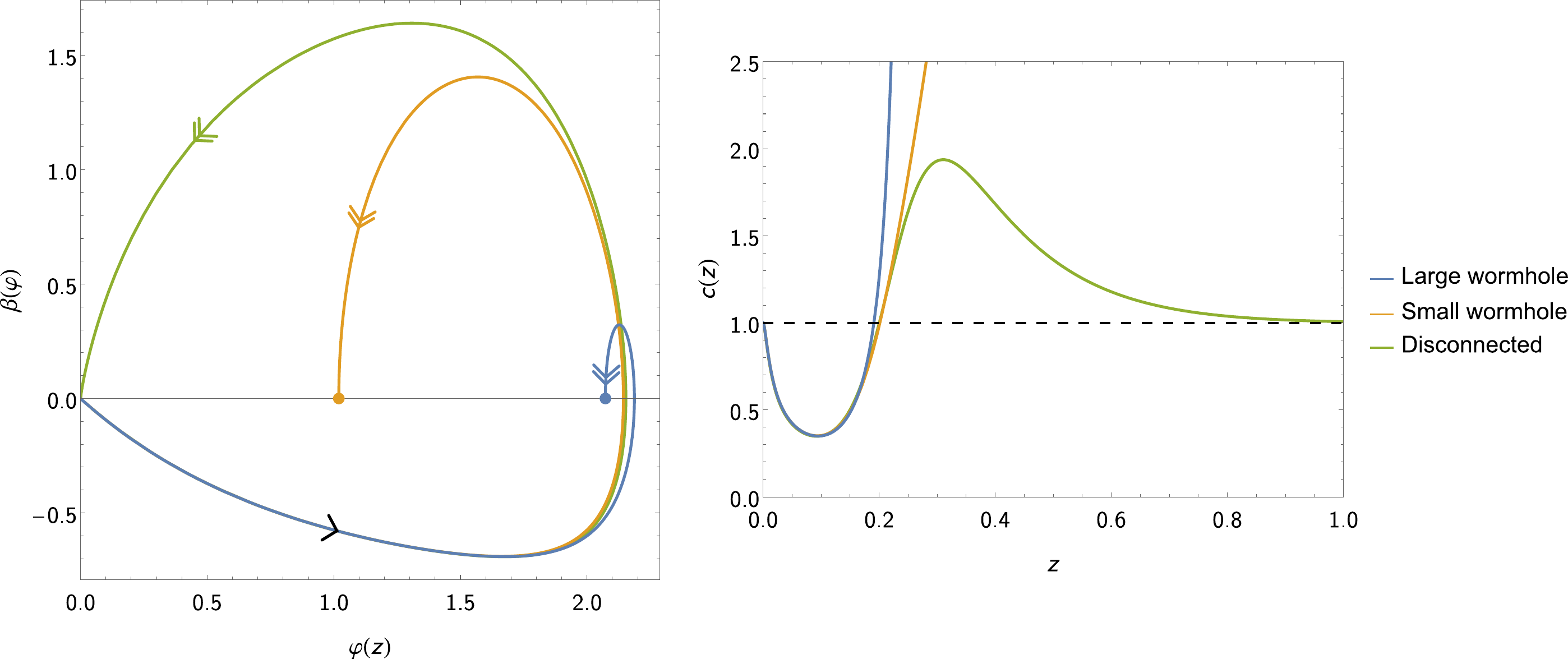}
    \caption{RG flow functions for the large wormhole, small wormhole, and disconnected solution for $d=3,~\Delta=2,~J/k=2$. \textit{Left.} The two-branched $\beta$-function. Arrows indicate the direction of flow along each branch: the black single arrow highlights the shared near-UV trajectory on the $\beta < 0$ branch, while double arrows label the differing behaviors on the $\beta > 0$ branch. Solid circles show where the flow terminates for wormholes. \textit{Right.} The holographic $c$-function. $c(z)$ diverges for wormholes as the flow reaches the throat.}
    \label{fig:all_3_RG}
\end{figure}

Having obtained $A(z)$ and $\varphi(z)$ for all three saddles, we can now study how the holographic flow functions for the folded wormholes compare with their disconnected cousin.
In terms of the FLRW scale factor in FG coordinates $a(z)$, they are
\begin{equation} \label{eq:beta_c_func_FRW}
    \beta(\varphi) = -\frac{\dd \varphi(z)}{\dd \ln z}, \qquad c(z) = c_0 \left(-z\frac{a'(z)}{a(z)} \right)^{1-d},
\end{equation}
where for the reader's convenience we have included the $\beta$-function which is unchanged from before.
The flow functions for all three saddles are plotted together in Fig.~\ref{fig:all_3_RG} for comparison.
Like the boomerang flows of Section~\ref{ssec:boomerang}, the wormhole flows also have multivalued $\beta$-functions with one positive and one negative branch.
Both small and large wormholes share their near-UV trajectory with the boomerang flow before differing in the IR.
Instead of returning to the origin, after switching branches at $\varphi_\text{max}$ all wormhole flows end at the throat, $\varphi_0$ defined in Eq.~\eqref{eq:FRW_IC1}.
This is how the ``pinching off'' of the $S^0$ in $A(z)$ manifests itself in the RG perspective.
A crucial observation is that $\varphi_0$ is again a turning point of the RG flow, not a fixed point; the RG acceleration in Eq.~\eqref{eq:RG_accel} is finite and negative at $\varphi_0$.\footnote{In fact, if we extend the RG flow by working with the unfolded wormhole, we will find that $\varphi_0$ is another branch point where the $\beta$-function continues past the throat to another negative branch.}

How do we see that the IR theory is ``gapped''?
In the conformal FLRW metric of Eq.~\eqref{eq:conformal_FRW}, any $\mathbb{Z}_2$-symmetric wormhole has finite range in the $\eta$ direction.
It follows from \cite{Antonini:2022opp} that the corresponding Sturm--Liouville problem for a probe scalar field on the wormhole background has a discrete spectrum of eigenvalues.
If we were working in Lorentzian signature, we could immediately conclude that this implies a discrete particle spectrum for the dual CFT and call it a day.
But because we have a Euclidean wormhole, we need to work a little harder; the Euclidean theory does not come equipped with a Hilbert space a priori.\footnote{One has to specify a Euclidean ``time'' direction to define a Hamiltonian operator and construct a Hilbert space, provided the Osterwalder--Schrader axioms \cite{Osterwalder:1974tc} are satisfied. The Hilbert space of the Lorentzian quantum field theory after analytic continuation therefore depends on how the Euclidean theory is time-sliced.}

The AdS/CFT extrapolate dictionary tells us that two-point correlators of CFT operators can be obtained by extrapolating bulk vacuum correlators to the boundaries at $\pm \eta_*$ \cite{Gubser:1998bc, Witten:1998qj, Freedman:1998tz}.
Denoting an insertion of the operator $\mathcal{O}(\xvec)$ on the left boundary at $-\eta_*$ by $\mathcal{O}_-(\xvec)$ and an insertion on the right boundary at $+\eta_*$ by $\mathcal{O}_+(\xvec)$, for $d=3$ the two-point functions are (see Section~2.1 of \cite{Antonini:2022opp} for details)
\begin{equation} \label{eq:same_correlator}
    \braket{\mathcal{O}_\pm(\xvec)\mathcal{O}_\pm(\yvec)} = \frac{1}{4\pi |\xvec-\yvec|} \sum_{n=0}^\infty u_n^2 \exp \left(-\sqrt{\lambda_n} |\xvec-\yvec|\right),
\end{equation}
for same-boundary insertions and 
\begin{equation} \label{eq:cross_correlator}
    \braket{\mathcal{O}_\pm(\xvec)\mathcal{O}_\mp(\yvec)} = \frac{1}{4\pi |\xvec-\yvec|} \sum_{n=0}^\infty (-1)^n u_n^2 \exp \left(-\sqrt{\lambda_n} |\xvec-\yvec|\right),
\end{equation}
for cross-boundary insertions.
In the above expressions, the mode sums are over the normalizable basis of solutions for the Klein--Gordon equation on the wormhole background.\footnote{By a normalizable basis, we mean in the same sense as standard quantization on an AdS background, e.g. as in Eq.~\eqref{eq:extrap_coeffs_example}. These are $L^2$-normalizable solutions to the Klein--Gordon equation.}
Given a wormhole scale factor $a(\eta)$, the coefficients $u_n$ of the normalizable modes and the eigenvalues $\lambda_n$ can always be determined by solving the corresponding Sturm--Liouville problem \cite{Antonini:2022opp}.
As $|\xvec - \yvec| \to \infty$, the leading contribution for both types of correlators come from the $n=0$ term
\begin{equation}
    \braket{\mathcal{O}_+(\xvec)\mathcal{O}_\pm(\yvec)} \sim \frac{u_0^2}{4\pi |\xvec - \yvec|} \exp\left(-\sqrt{\lambda_0} |\xvec - \yvec|\right),
\end{equation}
and similarly for $\braket{\mathcal{O}_-(\xvec)\mathcal{O}_\mp(\yvec)}$.
This exponential falloff of a two-point function is the hallmark of a gapped theory. In Euclidean signature, this implies that the correlation length of the system is finite and determined by the lowest eigenvalue, $\xi=1/\sqrt{\lambda_0}$, i.e. the ``mass gap''.

\begin{figure}[h]
    \centering
    \includegraphics[width=0.6\linewidth]{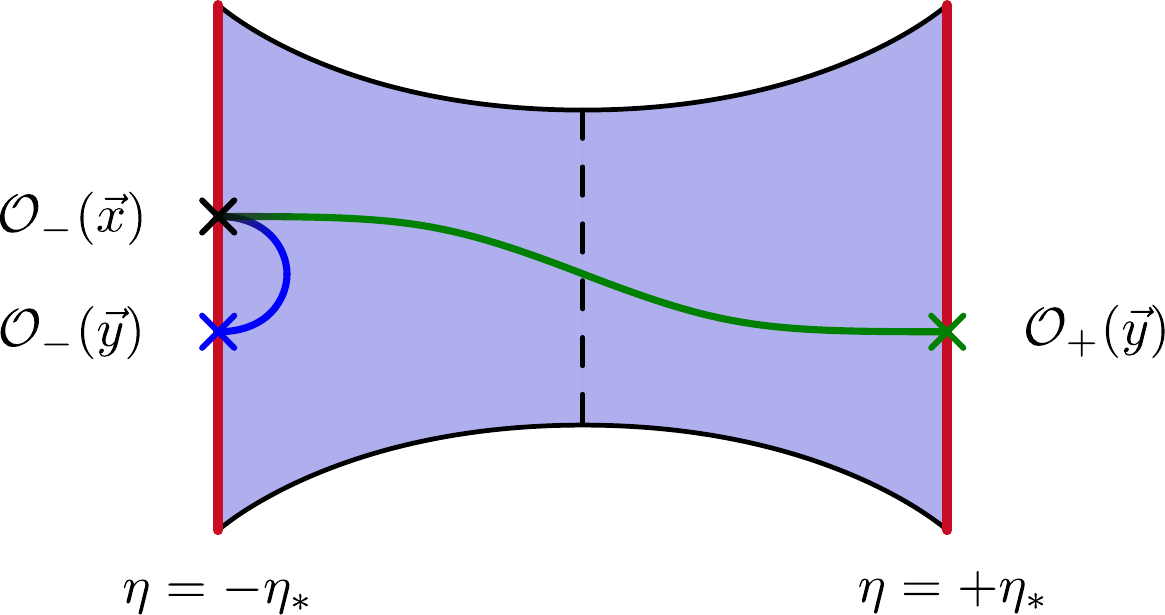}
    \caption{Bulk representation of the two types of boundary two-point functions for two CFTs connected via a bulk wormhole. Given an operator insertion, e.g. $\mathcal{O}_-(\xvec)$ on the left boundary (black cross), we can have either another insertion on the same boundary $\mathcal{O}_-(\yvec)$ (blue cross), giving the same-boundary correlator $\braket{\mathcal{O}_-(\xvec) \mathcal{O}_-(\yvec)}$ shown by the blue line, or an insertion on the opposite boundary $\mathcal{O}_+(\yvec)$ (green cross), giving the cross-boundary correlator $\braket{\mathcal{O}_-(\xvec) \mathcal{O}_+(\yvec)}$ shown by the green line. As $\yvec$ approaches $\xvec$, the blue line contracts to a point and $\braket{\mathcal{O}_-\mathcal{O}_-}$ diverges, whereas the green line continues to thread the wormhole throat and $\braket{\mathcal{O}_- \mathcal{O}_+}$ remains finite.}
    \label{fig:extrap_correlators}
\end{figure}

Can the presence of the wormhole throat be detected in the UV?
A geometric argument illustrated in Fig.~\ref{fig:extrap_correlators} highlights the fact that the pinching off of the $S^0$ and thus the gapping of the boundary field theory is strictly an IR phenomenon.
As $|\xvec - \yvec| \to 0$, scalar two-point functions in standard quantum field theory diverge as $|\xvec - \yvec|^{-2\Delta}$.
This is the leading UV behavior that one would expect from the same-boundary correlators in Eq.~\eqref{eq:same_correlator}.
The story for cross-boundary correlators is somewhat more subtle.\footnote{The absence of short-distance singularities in wormhole cross-correlators, and more generally the analytic structure of wormhole two-point functions, were first studied in \cite{Betzios:2019rds} and further developed in \cite{Betzios:2021fnm}.} 
For two disconnected boundaries, the cross-correlator trivially vanishes; the two boundary theories do not talk to each other.
For two boundaries connected via a bulk wormhole, however, not only are cross-correlators nonzero, but they also remain finite as $\xvec$ approaches $\yvec$.
From the perspective of the folded wormhole, this is due to the fact that the operator insertions are on opposite points of $S^0$, so we can have $\xvec = \yvec$ without insertions being physically on top of each other.
Because the UV divergence of same-boundary correlators drowns out any finite IR contribution from the wormhole geometry, same-boundary correlators alone are insufficient to probe bulk connectedness.\footnote{Another way to understand this is that there is always a unique bulk geodesic connecting two points on the same boundary. Such a geodesic extends into the wormhole, but never passes the throat. From this point of view, same-boundary correlators are inherently incapable of detecting the presence of the wormhole throat.}
Only cross-correlations or large separations can ``see'' the wormhole throat; bulk connectedness is invisible in the UV.



\section{Which saddle dominates the gravitational path integral?} \label{sec:saddles}

We have seen from Sections~\ref{ssec:numeric_sols_disc} and \ref{ssec:param_space} that above a critical threshold for the scalar source, three distinct geometries solve the Einstein-scalar equations of motion with two boundaries. 
There is one disconnected solution consisting of two disjoint copies of the one-boundary geometry found in Section~\ref{sec:1bdy}, and two connected solutions, namely the small and large wormholes found in Section~\ref{sec:wormholes}.
This raises the question of which geometry dominates the semiclassical approximation to the gravitational path integral with two boundaries.
In this section, we find the answer by calculating and comparing the renormalized on-shell action for each saddle to determine their dominance in parameter space.
We will see that unlike the models studied in \cite{Marolf:2021kjc}, the large wormhole dominates whenever it exists and that a Hawking--Page-like transition between the disconnected and connected phases is absent.
We defer the discussion of its implications on the problem of factorization in holography to Section~\ref{sec:discussion}.

Let us briefly recall the holographic renormalization procedure.
The Euclidean action as introduced in Eq.~\eqref{eq:bare_S} is generically a divergent quantity when evaluated on saddles with AdS asymptotics.
To obtain physically meaningful values for the Euclidean action, we first regularize the action by integrating it up to a cutoff surface near the conformal boundary.
In FG coordinates, this is the hypersurface defined by $z = \epsilon$.
Once the action has been regularized, we must then include the appropriate local counterterms at $z = \epsilon$ to render the on-shell Euclidean action finite as $\epsilon \to 0$ \cite{deHaro:2000vlm, Skenderis:2002wp}.
The complete renormalized on-shell action is therefore schematically written as
\begin{equation} \label{eq:S_ren}
    S_{\text{ren}} = \lim_{\epsilon \to 0} \Big(S_{\text{EH}} + S_{\text{GHY}} + S_{\Phi} + S_{\text{ct}} \Big).
\end{equation}
When correctly renormalized, the final expression for $S_\text{ren}$ should not depend on the location of the cutoff surface, $z=\epsilon$.
In our case where $d=3, \Delta=2$, taking into account gravitational backreaction due to the scalar fields, the holographic counterterms are (see Appendix~\ref{app:counterterms} for a detailed derivation)
\begin{equation} \label{eq:S_ct}
    S_{\text{ct}} = \frac{2}{\kappa} \int_{z=\epsilon} d^3x \sqrt{h} + \frac{1}{2\kappa} \int_{z=\epsilon} d^3x \sqrt{h}~R_h + \frac{1}{4\kappa} \sum_{I=1}^3 \int_{z=\epsilon} d^3x \sqrt{h}~ \Phi^*_I \Phi_I,
\end{equation}
where $h_{ij}$ is the induced metric on $z=\epsilon$, and $R_h$ is its scalar curvature.
More generally, the terms appearing in $S_{\text{ct}}$ will differ depending on the boundary dimension $d$ and the operator dimension $\Delta$.
We are now ready to evaluate the on-shell action for the three saddles.

\subsection{On-shell action for the disconnected saddle}

We first calculate the renormalized Euclidean action for the one-boundary geometry of Section~\ref{sec:1bdy}.
Evaluating the canonical action Eq.~\eqref{eq:bare_S} on the sinusoidal scalar ansatz in FG coordinates given by Eqs.~\eqref{eq:scalar_ansatz} and \eqref{eq:FG_ansatz}, we get
\begin{align}
    S_{\text{EH}}^{\text{(FG)}} &= \frac{3}{2\kappa} \int d^3x \int_{\epsilon}^{\infty} dz ~\frac{A^{3/2}(z)}{z^2} \left(\frac{A''(z)}{A(z)} - \frac{3}{z} \frac{A'(z)}{A(z)} + \frac{2}{z^2} \right), \label{eq:S_EH_FG} \\
    S_{\text{GHY}}^{\text{(FG)}} &=  -\frac{3}{\kappa} \int d^3x~ \frac{A^{3/2}(\epsilon)}{\epsilon^3}\left( 1 - \frac{\epsilon}{2}\frac{A'(\epsilon)}{A(\epsilon)} \right), \\
    S_{\Phi}^{\text{(FG)}} &= \frac{3}{4\kappa}\int d^3x \int_{\epsilon}^{\infty} dz~ \frac{A^{3/2}(z)}{z^2} \left( \varphi'^2(z) + \frac{k^2}{A(z)}  \varphi^2(z) + \frac{m^2}{z^2} \varphi^2(z) \right).
\end{align}
The counterterm action is
\begin{equation}
    S_{\text{ct}}^{\text{(FG)}} = \frac{1}{\kappa}\int d^3 x~\frac{A^{3/2}(\epsilon)}{\epsilon^3} \left( 2 + \frac{3}{4} \varphi^2(\epsilon) \right).
\end{equation}
Note that the curvature counterterm from Eq.~\eqref{eq:S_ct} vanishes identically since the boundary metric is flat.
Because none of the integrals above depend on $\xvec$, we can define $\int \dd^3x \equiv \Vol(\mathbb{R}^3)$ and factor out the formally infinite integral over $\mathbb{R}^3$ from Eq.~\eqref{eq:S_ren} to work with the renormalized action density $s_{\text{ren}}^{\text{(FG)}}$ instead.\footnote{In the following, we will refer to the \emph{action density} simply as the \emph{action}. Its precise meaning should hopefully be clear from context.}

At this point, we can insert the numerical solutions for $A(z)$ and $\varphi(z)$ found in Section~\ref{ssec:numeric_sols_disc} to obtain $s_{\text{ren}}^{\text{(FG)}}(\widetilde{J})$ for particular solutions.
We emphasize that because one-boundary solutions are parametrized by the dimensionless source strength $\widetilde{J} \equiv J/k^{d-\Delta}$ (or equivalently, $J$ and $k$), the renormalized action is ultimately a function of $\widetilde{J}$.
To compute the remaining $z$ integrals, in practice we will numerically integrate between $\epsilon < z < z_{\text{max}}$.
This is sensible to do since we have already seen from Eq.~\eqref{eq:z_infty_def} that the metric quickly relaxes to Poincar\'e-AdS in the bulk.
So long as we choose $z_{\text{max}}$ such that the behavior in Eq.~\eqref{eq:z_infty_def} persists, we can analytically estimate the truncated part of the action (density) integral to be just the last term of Eq.~\eqref{eq:S_EH_FG},
\begin{equation}
    s_{\text{error}} = \frac{3}{\kappa} \int_{z_{\text{max}}}^\infty dz~ \frac{A_{\infty}}{z^4} = \frac{1}{\kappa} \frac{A_{\infty}}{z_{\text{max}}^3}.
\end{equation}
Since $0 < A_\infty < 1$ always, even for seemingly small values like $z_{\text{max}}=2$, the error is at most $s_{\text{error}} < 1/4$, with typical values of $A_\infty \sim 0.7$ giving  $s_{\text{error}} \sim 0.18$.
As the on-shell action itself is at least $O(10^3)$, the error due to truncating the integral at finite $z$ is negligible.
Given a choice of $\widetilde{J}$, once we combine all the components of $s_{\text{ren}}^{\text{(FG)}}$ for a single copy of the one-boundary geometry, we can simply double it to get the full action for the disconnected saddle, $s^\text{(disc.)}(\widetilde{J}) \equiv 2s_{\text{ren}}^{\text{(FG)}} (\widetilde{J})$.

\subsection{On-shell action for connected saddles}

Next we turn to the wormhole geometries of Section~\ref{sec:wormholes}.
Working with the sinusoidal scalar ansatz in conformal FLRW coordinates given by Eqs.~\eqref{eq:scalar_ansatz} and \eqref{eq:conformal_FRW}, each component of the canonical action Eq.~\eqref{eq:bare_S} is given by
\begin{align}
    S_{\text{EH}}^\text{(conn.)} &= \frac{3}{\kappa} \int d^3x \int_{-\eta_*}^{\eta_*} d \eta ~\Big(\ddot{a}(\eta) a(\eta) - a^4(\eta)\Big), \\
    S_{\text{GHY}}^\text{(conn.)} &=  -\frac{3}{\kappa} \int d^3x~ \dot{a}(\eta_*) a(\eta_*) +\frac{3}{\kappa} \int d^3x~ \dot{a}(-\eta_*) a(-\eta_*), \\
    S_{\Phi}^\text{(conn.)} &= \frac{3}{4\kappa}\int d^3x \int_{-\eta_*}^{\eta_*} d \eta~ \Big( \dot{\varphi}^2(\eta) a^2(\eta) + k^2 a^2(\eta) \varphi^2(\eta) + m^2 a^4(\eta) \varphi^2(\eta) \Big).
\end{align}
The two terms in $S_{\text{GHY}}^\text{(conn.)}$ have opposite sign due to the outward normal pointing in opposite directions. 
The corresponding holographic counterterms are (one set for each boundary),
\begin{equation}
\begin{split}
    S_{\text{ct}}^\text{(conn.)} &= \frac{2}{\kappa}\int d^3 x~a^3(\eta_*) + \frac{3}{4\kappa} \int d^3x~a^3(\eta_*) \varphi^2(\eta_*) \\
    &+\frac{2}{\kappa}\int d^3 x~a^3(-\eta_*) + \frac{3}{4\kappa} \int d^3x~a^3(-\eta_*) \varphi^2(-\eta_*).
\end{split}
\end{equation}
Because our wormholes are $\mathbb{Z}_2$-symmetric, $a(\eta)$ and $\varphi(\eta)$ are even functions, so
\begin{equation}
    a(\eta) = a(-\eta), \qquad \dot{a}(\eta) = -\dot{a} (-\eta), \qquad \ddot{a}(\eta) = \ddot{a}(-\eta).
\end{equation}
The same is true of $\varphi$ and its $\eta$-derivatives.
This means the on-shell action above can be simplified into the following form for numerics
\begin{align}
    S_{\text{EH}}^\text{(conn.)} &= \frac{6}{\kappa} \int d^3x \int_0^{\eta_*} d \eta ~\Big(\ddot{a}(\eta) a(\eta) - a^4(\eta)\Big), \label{eq:S_EH_conn} \\
    S_{\text{GHY}}^\text{(conn.)} &=  -\frac{6}{\kappa} \int d^3x~ \dot{a}(\eta_*) a(\eta_*), \\
    S_{\Phi}^\text{(conn.)} &= \frac{3}{2\kappa}\int d^3x \int_0^{\eta_*} d \eta~ \Big( \dot{\varphi}^2(\eta) a^2(\eta) + k^2 a^2(\eta) \varphi^2(\eta) + m^2 a^4(\eta) \varphi^2(\eta) \Big), \\
    S_{\text{ct}}^\text{(conn.)} &= \frac{4}{\kappa}\int d^3 x~a^3(\eta_*) + \frac{3}{2\kappa}\int d^3x~a^3(\eta_*) \varphi^2(\eta_*). \label{eq:S_ct_conn}
\end{align}
We can equivalently view this as calculating the Euclidean action for the folded wormhole.
Once again, factoring out $\Vol(\mathbb{R}^3)$ from the expressions above, we get the connected action density $s^\text{(conn.)}$ after summing all components together.
Unlike $s^\text{(disc.)}$, no truncation of the $z$-integral is necessary here since the entire wormhole is captured within a finite $\eta$-interval.
From Section~\ref{ssec:param_space}, we know there is one small and one large wormhole solution for a given $\widetilde{J} > \widetilde{J}_\text{crit}$.
When substituting numerical solutions for $a(\eta)$ and $\varphi(\eta)$ into Eqs.~\eqref{eq:S_EH_conn} to \eqref{eq:S_ct_conn}, there are therefore two categories of $s^\text{(conn.)}$ to compute: the small wormhole action $s^\text{(small)}(\widetilde{J})$, and the large wormhole action $s^\text{(large)}(\widetilde{J})$.

\subsection{Saddle dominance}
\label{ssec:saddle_dominance}

Let us precisely define the quantity we are trying to compute.
The two-boundary path integral for Einstein-scalar gravity can be formally written as
\begin{equation} \label{eq:2bdy_GPI}
    Z_\text{grav}[\partial \mathcal{M}] = \int_{\partial \mathcal{M}} \mathcal{D}g \mathcal{D}\Phi ~e^{-S[g, \Phi]},
\end{equation}
where $\partial \mathcal{M} \equiv (\Sigma_1, J_1) \sqcup (\Sigma_2,J_2)$ denotes the disjoint union of boundary manifolds and their respective scalar sources.
Since we are only interested in the case where both boundaries are $\mathbb{R}^d$ and have the same sinusoidal sources, we can set as boundary conditions
\begin{equation}
    \partial \mathcal{M} = (\mathbb{R}^d, \{J_I\}) \sqcup (\mathbb{R}^d, \{J_I\}),
\end{equation}
where $\{J_I\}$ denotes the collection of sinusoidal sources Eq.~\eqref{eq:sinusoidal_source}.
These sources are parametrized by the amplitude $J$ and the wavenumber $k$, or equivalently the dimensionless source strength $\widetilde{J} = J/k^{d-\Delta}$.
In the semiclassical limit where $G_N \to 0$, the leading contribution to Eq.~\eqref{eq:2bdy_GPI} is given by a sum over on-shell solutions to the equations of motion, or equivalently the saddle-point configurations of the action.
In the regime where $\widetilde{J} > \widetilde{J}_\text{crit}$, there are three competing saddles, so the path integral is
\begin{equation} \label{eq:saddle_point_approx}
    Z_\text{grav} \left[ \partial \mathcal{M} \right] \simeq \sum_{\text{saddle}} Z_\text{grav}^\text{(saddle)} \left[ \partial \mathcal{M} \right],
\end{equation}
where the ``saddle'' sum is over the disconnected geometry denoted ``disc.'', and the two wormhole geometries collectively denoted by ``conn.''.
The contribution of an individual saddle to the sum is written
\begin{equation}
    Z_\text{grav}^\text{(saddle)} \left[ \partial \mathcal{M} \right] = \exp \left(-S^\text{(saddle)}( \widetilde{J}) + \cdots \right),
\end{equation}
where the classical action is $O(G_N^{-1})$, and the dots denote one-loop and higher order corrections of $O(G_N^0)$ and beyond.
Each action term can be formally factored into an action density and the volume of the Euclidean $d$-plane, $S(\widetilde{J})=s(\widetilde{J})\Vol(\mathbb{R}^d)$.
To see which saddle dominates the path integral, we would like to compute the difference between action densities
\begin{equation} \label{eq:action_diff}
    \Delta s(\widetilde{J}) \equiv s^{\text{(disc.)}}(\widetilde{J}) - s^{\text{(conn.)}} (\widetilde{J}),
\end{equation}
where $s^{\text{(conn.)}}(\widetilde{J})$ is either the small or large wormhole action.
As Eq.~\eqref{eq:saddle_point_approx} shows, greater actions are exponentially suppressed relative to lesser actions.\footnote{Because the one-loop determinant is $O(G_N^0)$, this correction to the action of the dominant saddle is still exponentially larger than the classical action for subdominant saddles.}
In particular, $\Delta s > 0$ indicates that a disconnected saddle is subleading relative to the connected saddle by a factor of $\exp\left(-\Delta s \Vol(\mathbb{R}^d)\right)$, while $\Delta s < 0$ shows that a connected saddle is subleading by a factor of $\exp\left(+\Delta s \Vol(\mathbb{R}^d)\right)$.
Whether the path integral is in the connected or disconnected phase for a given $\widetilde{J}$ can therefore be determined by calculating $\Delta s(\widetilde{J})$.

In the same manner as Section~\ref{ssec:param_space}, we sample points for the action difference by extracting the source corresponding to a wormhole solution \`a la Eq.~\eqref{eq:extrap_coeffs_example}.
This is to ensure that connected and disconnected saddles can be compared on the same footing.
Our numerical workflow is then as follows: for each value of $J$, we choose a window of cutoff surfaces $\{\epsilon\}$ within which we perform the integrals in Eq.~\eqref{eq:action_diff} up to the cutoff surface, which we denote by $\Delta s(\widetilde{J}; \epsilon)$. 
Since divergent parts of the action density (namely the $1/\epsilon$ and $1/\epsilon^3$ terms) were removed by the addition of counterterms, any finite renormalized action density (and by extension any difference of them) is expected to behave like
\begin{equation} \label{eq:action_expansion}
    \Delta s(\epsilon) = \Delta s_0 + \Delta s_1 \epsilon + \Delta s_2 \epsilon^2 + \cdots,
\end{equation}
near the asymptotic boundary (see e.g. Eq.~\eqref{eq:S_ren_power_law}).
$\Delta s_0$ is then the desired value of the renormalized action density difference as $\epsilon \to 0$.
We find that the series expansion Eq.~\eqref{eq:action_expansion} constitutes a good fit for our values of $\Delta s(\widetilde{J}; \epsilon)$; an independent fit allowing for $1/\epsilon$ and $1/\epsilon^3$ terms gives coefficients that are consistent with zero.
Further, the value of $\Delta s_0$ extracted in this manner is insensitive to variations in the choice of $\epsilon$ window.
Consequently, we take this constant coefficient as the numerical approximation to the action density difference, Eq.~\eqref{eq:action_diff}.

\begin{figure}[h]
    \centering
    \includegraphics[width=0.75\linewidth]{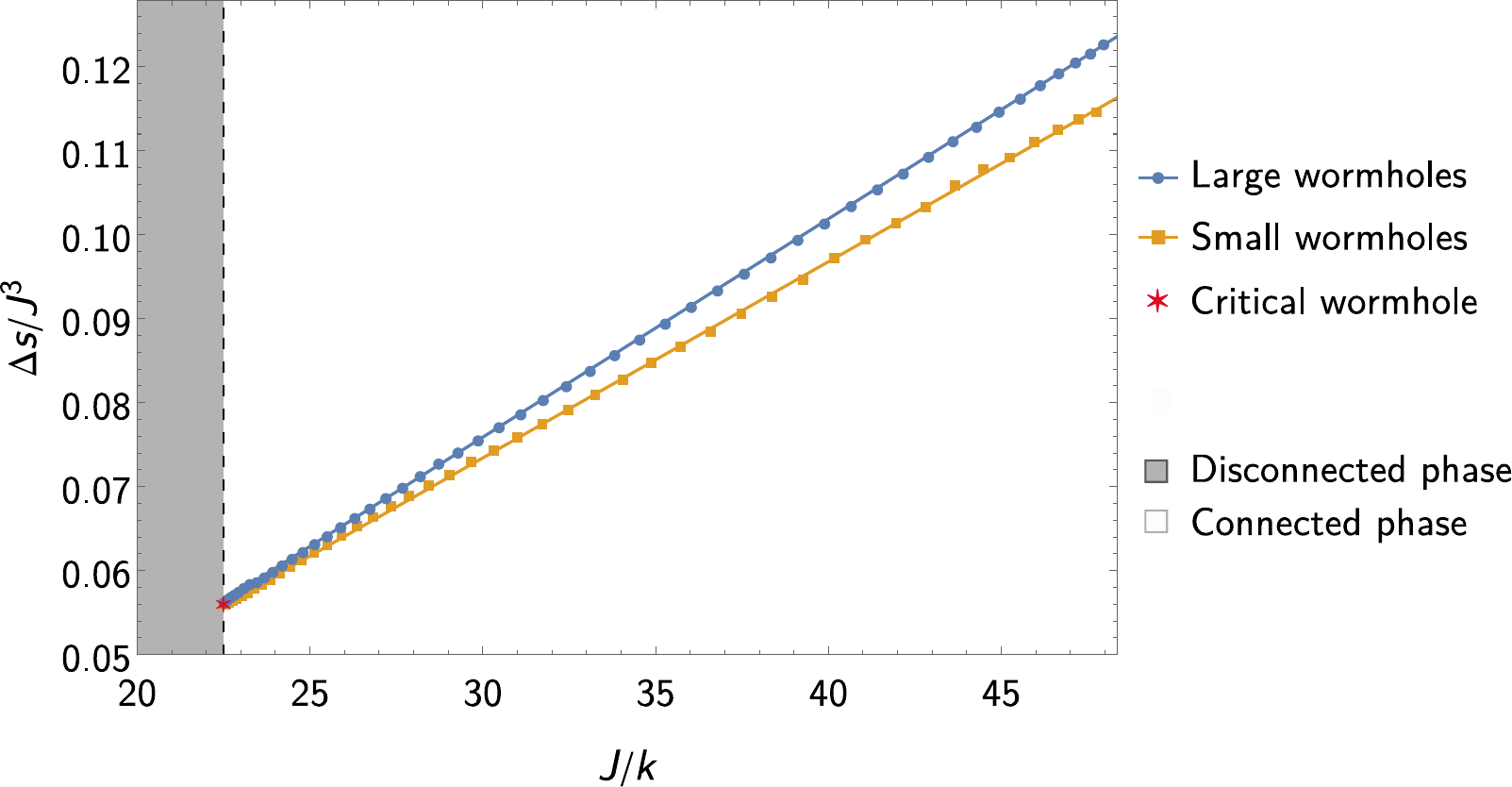}
    \caption{Phase diagram for the two-boundary gravitational path integral for $d=3,~\Delta=2$. Below a critical value of the source (dotted black line), only the disconnected saddle exists (grey region). Above this threshold, two connected saddles exist (white region). In particular, large wormholes (blue disks) always dominate over small wormholes (orange squares) and the disconnected saddle $(\Delta s/J^3 > 0)$. Away from the critical wormhole where both branches meet (red star), the linear fit suggests the action density difference behaves like $\Delta s \sim J^4/k$.}
    \label{fig:action_diff}
\end{figure}
Dimensional analysis tells us that the action density difference must take the form 
\begin{equation}
    \Delta s= J^{\frac{d}{d-\Delta}} f(\widetilde{J}) = J^3 f(J/k),
\end{equation}
where the second equality is specifically for the case of $d=3, \Delta = 2$.
To determine the unknown function $f(J/k)$, it is therefore illuminating to plot the dimensionless action density difference $\Delta s/J^3$ against the dimensionless source strength $J/k$.
In doing so, we obtain the main result of our paper: the gravitational phase diagram Fig.~\ref{fig:action_diff}.
We see that unlike the models studied in \cite{Marolf:2021kjc}, Fig.~\ref{fig:action_diff} shows that $\Delta s$ is positive and monotonically increasing for any size wormhole above $\widetilde{J}_\text{crit}$.
In fact, the curve for the large wormhole always lies above that of the small wormhole all the way down to critical wormhole where both branches coalesce into one point.
From this we can conclude that for $\widetilde{J}<\widetilde{J}_\text{crit}$, the disconnected geometry is the only saddle in the path integral, and the two boundaries never talk to each other, while for $\widetilde{J}\geq \widetilde{J}_\text{crit}$, the large wormhole is always the dominant saddle.\footnote{To be more precise, near the bifurcation point where the critical wormhole lives, the saddle-point approximation cannot clearly resolve the dominant saddle as the Euclidean actions for both wormhole branches are of comparable magnitude. Such a degeneracy may be lifted through a more careful treatment using e.g. Picard--Lefschetz theory \cite{Witten:2010cx, Witten:2021nzp}.} 
Crucially, there is no sign of a Hawking--Page-like transition; at the critical source strength $\widetilde{J} = \widetilde{J}_\text{crit}$, the action difference is already positive, indicating a sharp transition between the disconnected and connected phases of geometry.
The linear best-fit curve through our numerically obtained data points also suggests a simple dependence of the action density on the free parameters of the system: $f(J/k) \sim J/k$ or that $\Delta s \sim J^4/k$. 

\section{Discussion and open questions} \label{sec:discussion}

Let us take stock of the various results collected in this paper.
We started by studying semiclassical solutions to Einstein-scalar gravity assuming the sinusoidal scalar ansatz: a nontrivial but nonetheless tractable toy model of inhomogeneous matter capable of supporting homogeneous solutions in FLRW form.
We found two families of solutions satisfying our boundary conditions: one-boundary (disconnected) geometries and $\mathbb{Z}_2$-symmetric (connected) wormhole geometries.
The one-boundary geometry exists for any value of the source strength $\widetilde{J}$ and has a holographic RG interpretation of a boomerang flow where the field theory leaves the UV fixed point, and returns to the same CFT in the IR.
As for wormholes, we find that they exist only above a critical threshold for the source, and that there are generically two wormholes for any given source: a small and large wormhole.
These wormholes can also be interpreted in terms of holographic RG by folding them in half along the throat, in which case they represent gapped RG flows which terminate at the throat.
These two geometries naturally arise when studying the two-boundary gravitational path integral.
Most interestingly, comparing their on-shell actions reveal that whenever all three saddles exist, the large wormhole is always the dominant saddle; there is no Hawking--Page-like transition between the disconnected and connected phase of the path integral.
Below we will discuss the implications of our findings, and outline further avenues of inquiry for future work.

\subsection*{Microscopic realizations of boomerang and wormhole RG flows}

The boomerang flows we found for the disconnected solution in Section~\ref{ssec:boomerang} echo the earlier findings of \cite{Chesler:2013qla, Donos:2017ljs, Donos:2017sba}.
Indeed, \cite{Chesler:2013qla} studied an early version of boomerang-type flows by turning on a sinusoidal chemical potential in one of three boundary directions, obtaining an anisotropically rescaled AdS vacuum in the IR relative to the UV.
\cite{Donos:2017ljs, Donos:2017sba} later found a top-down and bottom-up example of true boomerang flows in Einstein-scalar theory with multiple AdS vacua, and found intermediate regimes of conformal invariance along the RG flow before a return to the original CFT in the IR.
These initial papers probed the tendency of their RG flows to return through holographic calculations of field theory observables such as the conductivity \cite{Chesler:2013qla}, refractive index \cite{Donos:2017ljs, Donos:2017sba}, and strip entanglement entropy \cite{Donos:2017sba}.
The disconnected geometry we found in Section~\ref{sec:1bdy} thus constitutes a minimal configuration that demonstrates boomerang behavior: an $\mathbb{R}^3$ boundary manifold, with only one AdS vacuum in the bulk, and a return to the same CFT (same AdS vacuum) in the IR.
By computing the $\beta$-function as a gradient flow of the bulk scalar field, we have not only recovered boomerang flows using a different approach, but also identified that our minimal boomerang flows are multivalued.
Because perturbations must decay for the bulk to return to the same vacuum, we expect that the $\beta$-functions of more general boomerang RG flows will have at least one turning point, and thus must also be multivalued.

It is important to stress that the boomerang and gapped RG flows dual to the disconnected and connected solutions are holographic descriptions derived from a semiclassical gravity calculation. 
To the best of our knowledge, there is no explicit field-theoretic realization of boomerang or gapped wormhole flows with multivalued $\beta$-functions in a strongly coupled theory.\footnote{For homogeneous relevant deformations, multivalued $\beta$-functions for exotic flows in holography have been catalogued extensively in \cite{Kiritsis:2016kog}.}
Given this, what can we infer about the boundary theory?
While the exotic RG flows themselves may not directly reveal details about the ensemble properties of the underlying microscopic theory, they can be a guide on other features the dual field theory should have.
For instance, the RG flows described by sinusoidal scalar wormhole geometries in Section~\ref{ssec:gapped} act as a nontrivial constraint on the UV and IR behavior of the putative field theory dual: the UV theory is a CFT on $\mathbb{R}^3 \times S^0$ which has to traverse a multivalued $\beta$-function before gapping out in the IR at an energy scale determined by the distance from the boundary to the wormhole throat.
Because the bulk wormhole leaves its imprint in the cross-correlator, a field-theoretic calculation should reproduce the same Euclidean ``mass gap'' structure as the extrapolate dictionary expression in Eq.~\eqref{eq:cross_correlator}. 
Additionally, since wormhole saddles only appear above a critical value of the source $\widetilde{J}_\text{crit}$, the gapped flows dual to wormhole geometries can only be probed in the non-perturbative regime of $\widetilde{J}$, i.e. for large deformations of the UV CFT. 
The phase diagram in Fig.~\ref{fig:action_diff} also raises the following question about our RG flows.
According to the results of Section~\ref{ssec:boomerang}, the disconnected saddle for the two-boundary gravitational path integral describes two independent boomerang RG flows for a CFT on $\mathbb{R}^3 \sqcup \mathbb{R}^3$.
The fact that the wormhole saddles dominate above $\widetilde{J}_\text{crit}$ then implies that the two independent flows join up in the bulk when the boundary source is large enough.
How do the RG flows of two seemingly independent, gapless CFTs merge to become a single RG flow that gaps in the IR? 
This is another facet of the factorization puzzle as viewed from the perspective of the holographic renormalization group.

\subsection*{Where is Hawking--Page?}

The bottom-up models of inhomogeneous wormholes studied by \cite{Marolf:2021kjc} generically exhibit HP-like transitions when the wormhole saddle becomes dominant; the action difference smoothly passes through zero, signaling a first-order phase transition.
In contrast, the phase diagram in Fig.~\ref{fig:action_diff} clearly shows no HP-like transition; the moment both wormhole branches spring into existence at $\widetilde{J}_\text{crit}$, the action difference is positive, and there is a discontinuity in the global minimum  of the action.
Phase transitions with a discontinuous jump in the free energy or Euclidean action are known as ``zeroth-order phase transitions'' and have been observed in superfluidity and superconductivity \cite{Maslov:2004}.
In the context of black hole thermodynamics, they have also appeared in reentrant phase transitions (large-small-large black hole transitions) of charged and rotating AdS black holes \cite{Gunasekaran:2012dq, Altamirano:2013ane, Kubiznak:2016qmn}.\footnote{See also \cite{Cong:2021jgb} for a different example of a zeroth-order phase transition in a holographic treatment of charged AdS black holes.}
Zeroth-order phase transitions are unusual in that they signal the sudden birth of a new, more thermodynamically favorable phase, rather than a competition between two existing phases.
Why do we not find a HP-like transition when the results of \cite{Marolf:2021kjc} suggest that we should find one? 
In using $\mathbb{R}^3$ as the boundary manifold to construct our sinusoidal scalar wormholes, we have one less length scale compared to \cite{Marolf:2021kjc}, associated with the compactness of either the $S^3$ or $T^3$ boundaries.
We suspect the lack of this extra length scale to be the reason why we do not see a HP-like phase transition, though the exact mechanism behind it is not clear.
To confirm this, it would be instructive to find the disconnected counterpart to the $T^3$ wormholes constructed in Appendix C of \cite{Marolf:2021kjc}, and check whether the action difference is smooth or discontinuous across the point in parameter space where saddles exchange dominance.

\subsection*{Implications for (non)-factorization}

The most striking feature of the phase diagram Fig.~\ref{fig:action_diff} is that whenever the wormhole saddles are available, they always have lower action than the disconnected geometry.
What can we learn about factorization with just the semiclassical gravity description?
Our phase diagram tells us that to leading order in the gravitational path integral, factorization is abruptly spoiled for a large enough boundary source, $\widetilde{J} \geq \widetilde{J}_\text{crit}$.
One obvious resolution of this apparent paradox would be to discount the single theory in favor of the ensemble interpretation.
Taking the ensemble seriously comes with several caveats, however.
It was already noted in \cite{Schlenker:2022dyo} that standard top-down examples of AdS/CFT, in particular ABJM theory, depends only on one integer $N$, and so it is not obvious as to what is being averaged over.\footnote{There are recent proposals on how to incorporate averaging into usual AdS/CFT examples in higher dimensions such as, e.g. averaging over the moduli space of $\mathcal{N}=4$ super-Yang--Mills \cite{Collier:2022emf}, or an ``asymptotic averaging'' over $N$ in ABJM theory \cite{Kudler-Flam:2026nzz}.}
Thus it is not clear to what extent we can expect the UV-completion of the sinusoidal scalar model to also have a suitable ensemble for averaging.
The putative ensemble would also have to fulfill some strange statistical properties.
In the framework of \cite{Coleman:1988cy, Giddings:1988cx, Giddings:1988wv} (made modern by \cite{Marolf:2020xie}), we can define the mean and variance of the statistical ensemble for the boundary partition function by $\braket{Z_\partial(\widetilde{J})} \simeq \exp\left(-S^\text{(disc.)}(\widetilde{J})/2 \right)$ and $\Var \big[ Z_\partial(\widetilde{J}) \big] \equiv \braket{Z_\partial^2 (\widetilde{J})} - \braket{Z_\partial (\widetilde{J})}^2$ respectively, where $\simeq$ indicates the leading contribution in the gravitational path integral.
In particular, only connected saddles contribute to the variance as the disconnected pieces from $\braket{Z_\partial}^2$ and $\braket{Z_\partial^2}$ cancel out exactly.
After normalizing, we see that our model requires fluctuations relative to the ensemble mean to behave like
\begin{equation}
    \frac{\Var \big[ Z_\partial(\widetilde{J}) \big]}{\braket{Z_\partial(\widetilde{J})}^2} \simeq 
    \begin{cases}
        0, & \widetilde{J} < \widetilde{J}_\text{crit} \\
        \exp \left( \Delta S(\widetilde{J})\right), & \widetilde{J} \geq \widetilde{J}_\text{crit}.
    \end{cases}
\end{equation}
Because no connected saddle exists below $\widetilde{J}_\text{crit}$, all partition functions in the ensemble are exactly self-averaging below $\widetilde{J}_\text{crit}$, i.e. the ensemble is $\delta$-function peaked as if there were just a single theory.
On the other hand, once the source reaches $\widetilde{J}_\text{crit}$, the ensemble is strongly non-self-averaging, and a typical member is not faithfully represented by the ensemble mean.
In particular, since $Z_\partial > 0$ by definition, exponentially large fluctuations (recall that $\Delta S>0$) implies that the ensemble must contain rare members whose partition functions are much larger than $\braket{Z_\partial}$, i.e. have much lower action than the disconnected saddle.

Similar statistical features have been previously observed in the wormhole dominated bottom-up models of \cite{Marolf:2021kjc}.
Because their wormholes dominate through a HP-like transition, compared to our model there is an intermediate regime between the wormholes existing and being dominant, $\widetilde{J}_\text{crit} < \widetilde{J} < \widetilde{J}_\text{HP}$.
In this regime where wormholes are subdominant, the boundary theory is still amenable to an ensemble interpretation as $\Var \big[ Z_\partial(\widetilde{J}) \big]$ is exponentially small, but still nonzero; the ensemble is sharply peaked but not a $\delta$-function.
This region of parameter space has been remarked in \cite{Marolf:2021kjc} as one that agrees with low-dimensional examples of ensembles in quantum gravity which are also sharply peaked (see e.g. \cite{Saad:2018bqo, Saad:2019lba, Stanford:2019vob, Marolf:2020xie}). 
It is therefore curious that our $\mathbb{R}^3$ sinusoidal wormhole model has no regime at all where non-factorization is exponentially suppressed.\footnote{Higher-loop corrections cannot change the fact that connected saddles do not exist below $\widetilde{J}_\text{crit}$, so any resolution must come from off-shell contributions like the constrained instantons of \cite{Cotler:2020lxj}.}

The above discussion rests on the assumption that our $\mathbb{R}^3$ wormholes do not suffer from instabilities.
It is therefore natural to question whether our wormhole saddles are perturbatively stable, i.e. whether their spectrum of quadratic fluctuations is free of negative modes.
While we did not carry out a stability analysis, there is a heuristic argument we can make as to why we expect at least the large branch of $\mathbb{R}^3$ wormholes to be stable.
The $T^3$ wormholes constructed in Appendix~C.2 of \cite{Marolf:2021kjc} have what appears to be the same, almost symmetric, large-small wormhole branch structure as our $\mathbb{R}^3$ wormholes (see Fig.~15 of \cite{Marolf:2021kjc}).
After an extensive stability analysis, the authors found no negative modes for the large branch of $T^3$ wormholes.
Their analysis also does not depend on the compactness of $T^3$, except for the fact that the wavenumber of the perturbations are quantized.
We thus anticipate the same analysis (with continuous perturbation wavenumbers) to carry over to the noncompact case, and expect the large branch of our $\mathbb{R}^3$ wormholes to also be stable to perturbative fluctuations.


\subsection*{Harmless speculation: integration detours and hidden saddles}

There are several possibilities for the boundary theory to make sense as a single theory.
This can be achieved if factorization is restored by some unknown mechanism.
Throughout this paper we have implicitly assumed that the integration contour for the gravitational path integral passes through the wormhole saddles, otherwise they would not contribute to the semiclassical sum over saddle points.
However, it is entirely possible that a more careful Picard--Lefschetz analysis will reveal that the integration contour does not actually pick up any of the wormhole saddles, so only the disconnected saddle remains valid.
Conversely, it is also possible that there are other contributions to the semiclassical approximation that are not captured geometrically by solutions to the Einstein-scalar equations of motion.
Non-perturbative objects like branes in a UV-completion, or non-geometric saddles such as the half-wormholes of \cite{Saad:2021rcu} or counter-wormholes of \cite{Gesteau:2024gzf} may conspire to cancel out the contribution from the large wormhole saddle and leave behind the disconnected saddle.
Let us try to estimate how big this mysterious contribution should be.
Suppose that the two-boundary partition function is schematically
\begin{equation}
    \braket{Z^2_\partial} \sim \exp\big(-S^\text{(disc.)}\big) + \exp\big(-S^\text{(conn.)}\big) + Z_?,
\end{equation}
where $Z_?$ denotes contributions from any hidden saddles.
For factorization to hold, clearly we must have $Z_? \approx - \exp\big(-S^\text{(conn.)}\big)$, in other words, the hidden saddles must be of the same order as the dominant wormhole saddle; it cannot be an exponentially small correction! 
The fact that $Z_?$ must be negative also suggests that perhaps it arises from some cancellation of erratic, complex phases, which is indeed what \cite{Saad:2021rcu} found for their half-wormholes.
If we take this estimate at face value, recovering factorization through this route then requires something drastically different from the semiclassical saddles we are familiar with.

\section*{Acknowledgments}
I am grateful to Mark Van Raamsdonk for his patience and guidance, especially for encouraging me to explore what happens when I fold my wormholes in half. 
I also thank Stefano Antonini, Chuanxin Cui, Kristan Jensen, Panos Betzios, and Alejandro Vilar L\'opez for helpful discussions and comments. 
I would like to extend my gratitude to the organizing committees of the QIQG 2025 conference held at the Perimeter Institute for Theoretical Physics, and the Extreme Universe 2025 workshop (YITP-T-25-01) held at the Yukawa Institute for Theoretical Physics, Kyoto University, where a preliminary version of this work was presented. 
I am supported by the Four Year Doctoral Fellowship (4YF) of the University of British Columbia.
\newpage

\bibliographystyle{jhep}
\bibliography{refs}

\appendix
\section{Holographic counterterms for $d=3, \Delta=2$} \label{app:counterterms}

Following the formalism developed in \cite{deHaro:2000vlm, Skenderis:2002wp}, we derive the holographic counterterms required to renormalize the on-shell action.
We start by rewriting the canonical action Eq.~\eqref{eq:bare_S} in trace-reduced form.
Recall that the Einstein equation in the bulk is
\begin{equation}
    R_{ab} - \frac{1}{2}Rg_{ab} + \Lambda g_{ab} = \kappa T_{ab},
\end{equation}
where the stress tensor for our scalar fields is
\begin{equation}
    T_{ab} = \frac{1}{2\kappa} \sum_{I=1}^d \left[ \partial_a \Phi^*_I \partial_b \Phi_I - \frac{1}{2}g_{ab} \Big(\partial_c \Phi^*_I \partial^c \Phi_I + m^2 \Phi^*_I \Phi_I \Big) \right].
\end{equation}
Taking the trace of the Einstein equation, and using the fact that $\Lambda = -d(d-1)/2$,
\begin{equation} \label{eq:tr_EE}
    R = \frac{1}{2} \sum_{I=1}^d \left( g^{ab} \partial_a \Phi^*_I \partial_b \Phi_I + \frac{d+1}{d-1} m^2 \Phi^*_I \Phi_I\right) - d(d+1).
\end{equation}
Plugging this into Eq.~\eqref{eq:bare_S}, we get the trace reduced action
\begin{equation} \label{eq:tr_red_s}
    S = \frac{d}{\kappa} \int_{\EM} d^{d+1}x \sqrt{g}  - \frac{1}{\kappa} \int_{\partial \EM} d^dx \sqrt{h} K - \frac{1}{2\kappa} \frac{m^2}{d-1} \int_{\EM} d^{d+1}x \sqrt{g}  \sum_{I=1}^d\Phi_I^*\Phi_I.
\end{equation}
We see that the canonical action has now been reduced to a bulk volume term, an unmodified Gibbons--Hawking--York (GHY) boundary term, and a scalar mass term.

To find the power law behavior of divergences near the boundary, it is standard to work in FG coordinates.
Once we determine the correct counterterms, they are coordinate independent so it will be straightforward to get explicit expressions for any choice of metric.
Starting with the gravitational part of the action, the most general FG metric takes the form
\begin{equation} 
    ds^2 = \frac{1}{z^2} \Big(dz^2 + \gamma_{ij}(z,x) dx^i dx^j \Big),
\end{equation}
where $\gamma_{ij}$ admits a power law expansion near the boundary located at $z=0$.
For our particular case where $d=3$, this asymptotic expansion is
\begin{equation} \label{eq:FG_expansion}
    \gamma_{ij} (z,x) = \gamma_{ij}^{(0)}(x) + z^2 \gamma_{ij}^{(2)}(x) + z^3 \gamma_{ij}^{(3)}(x) + O(z^4).
\end{equation}
From this, the determinant of $\gamma_{ij}$ can also be expanded as
\begin{equation}
    \gamma = \gamma^{(0)} \big(1 + z^2\tr \gamma^{(2)} + z^3 \tr \gamma^{(3)} + O(z^4) \big),
\end{equation}
where indices are contracted using $\gamma^{(0)}_{ij}$ so that e.g., $\tr \gamma^{(2)} = \gamma_{(0)}^{ij} \gamma^{(2)}_{ij}$.
Regularizing the trace-reduced action by implementing a cutoff surface at $z=\epsilon$, the bulk volume term becomes
\begin{equation}
    \frac{3}{\kappa} \int_{z\geq \epsilon} d^{4}x \sqrt{g} = \frac{1}{\kappa} \int_{z = \epsilon} d^3x \sqrt{\gamma^{(0)}} \left( \frac{1}{\epsilon^3} + \frac{3}{2\epsilon} \tr \gamma^{(2)} - \frac{3}{2} \ln\epsilon \tr \gamma^{(3)} + O(\epsilon) \right).
\end{equation}
While the logarithm may be alarming at first glance, this actually gets canceled by an equivalent piece of opposite sign coming from the scalar mass term, leaving only power law behavior behind.

To evaluate the GHY term, we need the induced metric on $z=\epsilon$
\begin{equation}
    ds^2 \Big|_{z=\epsilon} \equiv h_{ij} dx^i dx^j = \frac{1}{\epsilon^2} \gamma_{ij}(z,x) dx^i dx^j,
\end{equation}
so that its determinant is
\begin{equation}
    \sqrt{h} = \epsilon^{-3} \sqrt{\gamma^{(0)}} \left( 1 + \frac{\epsilon^2}{2} \tr \gamma^{(2)} + \frac{\epsilon^3}{2} \tr \gamma^{(3)} + O(\epsilon^4) \right).
\end{equation}
Choosing an outward pointing normal vector $n_a=(-1/z,~ 0)$, at the hypersurface  $z=\epsilon$,
\begin{equation}
    K_{ij} = \frac{1}{\epsilon^2} \gamma_{ij} - \frac{1}{2\epsilon} \partial_z \gamma_{ij}.
\end{equation}
Taking the trace with $h^{ij}$ to get the extrinsic curvature,
\begin{equation}
    K = 3 - \epsilon^2 \tr \gamma^{(2)} - \frac{3}{2} \epsilon^3 \tr \gamma^{(3)} + O(\epsilon^4).
\end{equation}
The GHY term is therefore
\begin{equation}
    -\frac{1}{\kappa} \int_{z=\epsilon} d^3x \sqrt{h}~K = -\frac{1}{\kappa} \int_{z=\epsilon} d^3x \sqrt{\gamma^{(0)}} \left( \frac{3}{\epsilon^3} + \frac{1}{2\epsilon} \tr \gamma^{(2)} + O(\epsilon) \right).
\end{equation}
Combining this with the volume term above, the full gravitational part of the trace-reduced action is
\begin{equation}
    S_{\text{grav}} = -\frac{1}{\kappa} \int_{z=\epsilon} d^3x \sqrt{\gamma^{(0)}} \left( \frac{2}{\epsilon^3} - \frac{1}{\epsilon} \tr \gamma^{(2)} + \frac{3}{2} \ln \epsilon \tr \gamma^{(3)} + O(\epsilon) \right).
\end{equation}

Next, let us look at the divergences coming from the scalar mass term.
Temporarily dropping the $I$ index on $\Phi_I$ to reduce notational clutter, for any $d$ and $\Delta$ with no matter anomaly (i.e., no logarithms), the scalar field has the following power law expansion near $z=0$
\begin{equation} \label{eq:phi_asymp}
    \Phi(z,x) = z^{d-\Delta} \big( \Phi_{(0)}(x) + z^2 \Phi_{(2)}(x) + \cdots \big) + z^\Delta \big( \Phi_{(2\Delta-d)}(x) + z^2 \Phi_{(2\Delta-d+2)}(x) + \cdots \big).
\end{equation}
In particular, $\Phi_{(0)}$ corresponds to the boundary source, and $\Phi_{(2\Delta-d)}$ corresponds to the VEV of the boundary scalar operator \cite{Gubser:1998bc, Witten:1998qj}.
In our case where $d=3, \Delta=2$, this is simply
\begin{equation} 
    \Phi(z,x) = \big(z \Phi_{(0)}(x) + z^3 \Phi_{(2)}(x) + \cdots \big) + \big(z^2 \Phi_{(1)}(x) + z^4 \Phi_{(3)}(x) + \cdots \big),
\end{equation}
keeping the ``source'' and ``VEV'' branches of the solution distinct.

To find the divergent powers coming from the scalar action, plug this all into the scalar mass term in Eq.~\eqref{eq:tr_red_s}.
For one copy of the scalar when $d=3, \Delta=2$, the result is\footnote{Hidden in the $O(\epsilon)$ terms are two terms that are finite for $\Delta = 2$, but are divergent for $5/2 < \Delta < 3$.}
\begin{equation}
    - \frac{m^2}{4\kappa} \int_{z \geq \epsilon} d^{4}x \sqrt{g}~ \Phi_I^*\Phi_I = \frac{1}{2\kappa} \int_{z=\epsilon} d^3x \sqrt{\gamma^{(0)}} \left( \frac{1}{\epsilon} \left|\Phi^I_{(0)}\right|^2 - 2\ln \epsilon ~\mathrm{Re} \left[\big(\Phi^{I}_{(0)}\big)^* \Phi^I_{(1)} \right] + O(\epsilon) \right)
\end{equation}
Going back to Eq.~\eqref{eq:tr_EE}, note that for arbitrary $d$ and $\Delta$ (again ignoring anomalies), expanding both sides in their respective power series in $z$, the $O(z^d)$ terms gives us the relation
\begin{equation}
    \tr \gamma^{(d)} = \frac{2 m^2}{d(d-1)} \sum_{I=1}^d \mathrm{Re} \left[  \big(\Phi^I_{(0)}\big)^* \Phi^I_{(2\Delta-d)} \right].
\end{equation}
For $d=3, \Delta=2$, this is just
\begin{equation}
    -\frac{3}{2}\tr \gamma^{(3)} = \sum_{I=1}^3 \mathrm{Re} \left[  \big(\Phi^I_{(0)}\big)^* \Phi^I_{(1)} \right].
\end{equation}
Thus summing over $I$, we find that
\begin{equation}
    - \frac{m^2}{4\kappa} \int_{z \geq \epsilon} d^{4}x \sqrt{g}~ \sum_{I=1}^3\Phi_I^*\Phi_I = \frac{1}{2\kappa} \int_{z=\epsilon} d^3x \sqrt{\gamma^{(0)}} \left( \frac{1}{\epsilon} \sum_{I=1}^3\left|\Phi^I_{(0)}\right|^2 + 3\ln \epsilon \tr \gamma^{(3)} + O(\epsilon) \right).
\end{equation}
The logarithm on the right-hand side exactly cancels out the logarithm coming from the gravitational part and we are left with just power law divergences as promised.
Combining the matter and gravitational parts of the trace-reduced action, we finally arrive at this expression for the regularized action
\begin{equation}
    S_{\text{reg}} = -\frac{1}{\kappa}\int_{z=\epsilon} d^3x \sqrt{\gamma^{(0)}} \left[ 2 \epsilon^{-3} - \left( \tr \gamma^{(2)} +  \frac{1}{2} \sum_{I=1}^3\left|\Phi^I_{(0)}\right|^2 \right) \epsilon^{-1} + O(\epsilon) \right].
\end{equation}

To obtain a coordinate invariant expression for the counterterms, we need to rewrite these divergent terms using boundary local quantities like $\Phi_I$ and scalars made from the induced metric $h_{ij}$.
First, note that the $O(z^2)$ terms in Eq.~\eqref{eq:tr_EE} gives us
\begin{equation}
    \tr \gamma^{(2)} = -\frac{1}{4}R^{(0)} - \frac{3}{8} \sum_{I=1}^3 \left| \Phi^I_{(0)} \right|^2,
\end{equation}                                           
where $R^{(0)}$ is the Ricci scalar of $\gamma^{(0)}_{ij}$.
The $\Phi_{(0)}$ term here is unique to $d=3, \Delta=2$ since the backreaction from the scalar shows up in the stress tensor at exactly $O\big(z^{2(d-\Delta)}\big)=O(z^2)$.
To leading order, we can then invert $\Phi_{(0)}, \gamma^{(0)}$, and $R^{(0)}$ like so
\begin{align}
    \Phi_{(0)} &= \epsilon^{-1} \big(\Phi + O(\epsilon) \big), \\
    R^{(0)} &= \epsilon^{-2} \big(R_h + O(\epsilon^2)\big), \\
    \sqrt{h} &= \sqrt{\gamma^{(0)}} \epsilon^{-3} \left(1 + \frac{\epsilon^2}{2} \tr \gamma^{(2)} + O(\epsilon^3) \right),
\end{align}
where $R_h$ is the Ricci scalar of $h_{ij}$.
Taking the holographic counterterms to have the opposite sign to $S_{\text{reg}}$ so that the divergences are canceled out, some harmless algebra results in
\begin{equation}
    S_{\text{ct}} = \frac{2}{\kappa} \int_{z=\epsilon} d^3x \sqrt{h} + \frac{1}{2\kappa} \int_{z=\epsilon} d^3x \sqrt{h}~R_h + \frac{1}{4\kappa} \sum_{I=1}^3 \int_{z=\epsilon} d^3x \sqrt{h}~ \Phi^*_I \Phi_I.
\end{equation}
The complete renormalized on-shell action for $d=3, \Delta=2$ is therefore
\begin{equation} 
    S_{\text{ren}} = S_{\text{EH}} + S_{\text{GHY}} + S_{\Phi} + S_{\text{ct}}.
\end{equation}
In our numerical analysis, we used the canonical form of the action given by Eq.~\eqref{eq:bare_S} as it turns out to be more numerically stable compared to the trace-reduced form of Eq.~\eqref{eq:tr_red_s}.
From the above analysis, we can also see that if the $\epsilon^{-3}$ and $\epsilon^{-1}$ divergences have been properly removed by the counterterms, the finite renormalized action should have a power law expansion near the boundary that behaves like
\begin{equation} \label{eq:S_ren_power_law}
    S_{\text{ren}}(\epsilon) = S_0 + S_1 \epsilon + S_2 \epsilon^2 + O(\epsilon^3).
\end{equation}
This provides a diagnostic for whether the action has been properly renormalized and is helpful when extracting the cutoff independent value of the action $S_0$.

\end{document}